\documentclass[]{nst} 

\usepackage{subfigure,dcolumn}
\usepackage{epstopdf}
\usepackage{mhchem}
\usepackage{upgreek}
\usepackage{float}
\usepackage{graphicx}
\usepackage{bm}
\usepackage{color}
\usepackage{CJKutf8}
\usepackage{xcolor}
\usepackage[normalem]{ulem}
\usepackage{orcidlink}

\usepackage{ragged2e}
\usepackage{wrapfig}
\usepackage{amsmath}

\graphicspath{{figs/}}  
\allowdisplaybreaks

\newcommand{\mev}{\mathrm{MeV}}
\newcommand{\tev}{\mathrm{TeV}}

\usepackage{tikz,xcolor,hyperref}

\definecolor{lime}{HTML}{A6CE39}
\DeclareRobustCommand{\orcidicon}{
	\begin{tikzpicture}
	\draw[lime, fill=lime] (0,0) 
	circle [radius=0.16] 
	node[white] {{\fontfamily{qag}\selectfont \tiny ID}};
	\draw[white, fill=white] (-0.0625,0.095) 
	circle [radius=0.007];
	\end{tikzpicture}
	\hspace{-2mm}
}
\foreach \x in {A, ..., Z}{%
	\expandafter\xdef\csname orcid\x\endcsname{\noexpand\href{https://orcid.org/\csname orcidauthor\x\endcsname}{\noexpand\orcidicon}}
}

\newcommand{\X}{X(3872)}

\newcommand{\FudanU}{\affiliation{Key Laboratory of Nuclear Physics and Ion-beam Application (MOE), Institute of Modern Physics, Fudan University, Shanghai 200433, China}}
\newcommand{\NSFCThC}{\affiliation{Shanghai Research Center for Theoretical Nuclear Physics, NSFC and Fudan University, Shanghai 200438, China}}

\newcommand{\THU}{\affiliation{Department of Physics, Tsinghua University, Beijing 100084, China}}

\newcommand{\itp}{\affiliation{CAS Key Laboratory of 
Theoretical Physics, Institute of Theoretical Physics,\\
Chinese Academy of Sciences, Beijing 100190, China}}

\newcommand{\ucas}{\affiliation{School of Physical Sciences, 
University of Chinese Academy of Sciences, Beijing 100049, 
China}}

\newcommand{\peng}{\affiliation{Peng Huanwu Collaborative 
Center for Research and Education, Beihang University, Beijing 
100191, China}}

\newcommand{\scnt}{\affiliation{Southern Center for Nuclear-Science Theory (SCNT), Institute of Modern Physics,\\ 
Chinese Academy of Sciences, Huizhou 516000, China}}

\newcommand{\moe}{\affiliation{Key Laboratory of Atomic and Subatomic Structure and Quantum Control (MOE), Guangdong Basic Research Center of Excellence for Structure and Fundamental Interactions of Matter, Institute of Quantum Matter, South China Normal University, Guangzhou 510006, China
}}

\newcommand{\iqm}{\affiliation{Guangdong-Hong Kong Joint Laboratory of Quantum Matter, Guangdong Provincial Key Laboratory of Nuclear Science, Southern Nuclear Science Computing Center, South China Normal University, Guangzhou 510006, China}}

\begin{document}

\title{Production of exotic hadrons in $pp$ and nuclear collisions}

\thanks{This review is dedicated to Professor Wenqing Shen on the occasion to celebrate his 80th birthday. All authors contributed equally to this work.
}

\author{Jinhui Chen\orcidlink{0000-0001-7032-771X}}
\FudanU\NSFCThC

\author{Feng-Kun Guo\orcidlink{0000-0002-2919-2064}}
\itp\ucas\peng\scnt

\author{Yu-Gang Ma\orcidlink{0000-0002-0233-9900}} 
\FudanU\NSFCThC

\author{Cheng-Ping Shen\orcidlink{}}
\FudanU\NSFCThC

\author{Qiye Shou\orcidlink{}}
\FudanU\NSFCThC

\author{Qian Wang\orcidlink{0000-0002-2447-111X}}
\iqm\moe\scnt

\author{Jia-Jun Wu\orcidlink{}}
\ucas

\author{Bing-Song Zou\orcidlink{}}
\THU \itp \peng \scnt


\begin{abstract}
Exotic hadrons beyond the conventional quark model have been discovered in the past two decades.
Investigations of these states can lead to deep understanding of nonperturbative dynamics of the strong interaction.
In this concise review, we focus on the productions of exotic hadrons in $pp$, $p\bar p$, and nuclear collisions.
Experimental observations of light nuclei and hypernuclei, as prototypes of hadronic molecules, in heavy ion collisions will also be briefly discussed. 
\end{abstract}


\maketitle


\section{Introduction}

Gell-Mann and Zweig introduced quark model to classify hadrons discussed thus far~\cite{Gell-Mann:1964ewy,Zweig:1964jf}.
Mesons and baryons were successfully classified as being made out of $q\bar q$, $qq\bar q\bar q$, etc. and $qqq$, $qqqq\bar q$, etc., respectively, and the lowest-lying states were supposed to be  quark-antiquark and three-quark states. 
Quantum chromodynamics (QCD) brought dynamics to the quark model, inspiring the construction of constituent quark models with various potentials among quarks and gluons (see, e.g., Refs.~\cite{Godfrey:1985xj, Capstick:1986ter, Glozman:1995fu}). 
In the consistent quark models, most of the hadrons observed before the year of 2003 could be successfully described in terms of $q\bar q$ mesons and $qqq$ baryons, with a few exceptions such as the lightest scalar mesons and the $J^P=1/2^-$ $N(1535)$ and $\Lambda(1405)$. 
Consequently, from the quark model point of view, the $q\bar q$ mesons and $qqq$ baryons were considered ordinary hadrons, while those with other possible color-singlet valence contents, defined as quarks and gluons responsible for the hadron quantum numbers, are generally called exotic hadrons.
These include multiquark states, hybrid mesons and baryons, and glueballs, while multiquark states can be further divided into compact multiquarks and hadronic molecules according to how the (anti)quarks are grouped together. 
For hadronic molecules, the quarks and antiquarks first form color-singlet hadrons, which then interact with each other through the residual strong force to form composite hadron systems; in this sense, hadronic molecules are not ``exotic'' but a rather natural extension of atomic nuclei to composite systems of other hadrons.

Study of how the colorful quarks and gluons are grouped inside hadrons is crucial in gaining new insights into the mechanism of color confinement.
Therefore, there have been decade-long efforts to search for hadrons with distinct exotic characteristics, in experiments including $e^+e^-$ annihilations, hadron collisions, electron-ion collisions and so on. Thanks to the high statistics of the new generations of experiments, many new hadron resonances have been discovered since 2003 with properties at odds with quark expectations of ordinary mesons and baryons, and they are good candidates for exotic hadrons. 
Most of them were mesonic states observed in the heavy quarkonium mass region, and are usually referred to collectively as $XYZ$ states.

These exotic hadron candidates have been searched for and measured in various experiments. Since QCD is intrinsically nonperturbative at low energies, it has been challenging to put these experimental observations into order to gain deep insights. Different theoretical methods have been used, such as lattice QCD, effective field theories, (unquenched) quark model, QCD sum rules and so on. Each one of them has advantages and drawbacks. So far a universal description of all these exotic hadron candidates is still out of reach. 
For recent reviews of experimental observations and theoretical investigations, see, e.g., Refs.~
\cite{Esposito:2016noz, Guo:2017jvc, Olsen:2017bmm,  Kalashnikova:2018vkv, Brambilla:2019esw, Guo:2019twa, Yang:2020atz, Chen:2022asf, Mai:2022eur, Liu:2024uxn}.

Here we will focus on the experimental observations of exotic hadrons in $pp$, $p\bar p$ and nuclear collisions, and discuss the productions of the exotic hadrons in these collisions. 
Productions of light nuclei and hypernuclei, as prototypes of hadronic molecules, in heavy ion collisions (HICs) will also be briefly discussed.

\section{Multiquark candidates with hidden charm and double charm} 
\label{sec:multiquarkexp}

Above the open-charm threshold, tens of unexpected states with properties inconsistent with expectations from the traditional quark model have been observed since 2003. A few similar states were also observed in the bottomonium mass region. Among them, the $Y$(mass) states have vector quantum numbers and show a strong coupling to hidden-charm or open-charm final states. The multiquark candidates, including tetraquark $Z_Q$(mass) and pentaquark  $P_Q$(mass),  with hidden $Q\bar Q$ pair are considered as explicitly exotic since they carry nonvanishing isospin and/or strangeness in addition to the $Q\bar Q$ pair. Here $Q$ refers to either charm or bottom quark.

The first hidden-charm state that triggered studies of the exotic states is the $X(3872)$, also known as $\chi_{c1}(3872)$, observed by the Belle experiment in 2003~\cite{Belle:2003nnu}. 
Besides the $X(3872)$, the $Z_c(4430)$ and $Z_b(10610/10650)$,
which are the first charged charmonium-like and bottomonium-like states with obvious exotic characteristics, were also observed by Belle~\cite{Belle:2007hrb,Belle:2011aa}. The first vector charmonium-like state $Y(4260)$ and pentaquark states $P_c(4450)$ and $P_c(4380)$ were observed by BaBar and LHCb experiments~\cite{BaBar:2005hhc,LHCb:2015yax}, respectively.

\subsection{Charmonium-like states}
\label{sec:charmoniumlike}

The $X(3872)$ was discovered by Belle in $B\to K \pi \pi J/\psi$ decays as a narrow peak in the invariant mass distribution of the  $ \pi \pi J/\psi$  final state~\cite{Belle:2003nnu}.
Ten years after its discovery, LHCb performed a full five-dimensional amplitude analysis in  $B^+ \to K^+ X(3872) \to K^+ \pi^+ \pi^- J/\psi$ decays
and unambiguously gave the $J^{PC}=1^{++}$ 
assignment~\cite{LHCb:2013kgk}. 

The most salient feature of the $\X$ is that its mass coincides almost exactly with the threshold of $D^0\bar D^{*0}$.
The mass and width of the $\X$ as given in the latest version of the Review of Particle Physics (RPP)~\cite{ParticleDataGroup:2024cfk} are $(3871.64 \pm 0.06)$~MeV and $(1.19\pm0.21)$~MeV, respectively.
Notice that these values were extracted using the Breit-Wigner (BW) parameterization. The BW parameterization works well only for isolated resonances far from thresholds of channels that the resonance can strongly couple to;
however, the $\X$ is different and thus the BW width reported in RPP is not a good approximation of twice the imaginary part of the $\X$ pole position in the complex energy plane.

Using the Flatt\'e parameterization that takes into account the nearby $D\bar D^*$ thresholds~\cite{Hanhart:2010wh}, the LHCb Collaboration reported the mass and width of the $\X$ as  $3871.69_{-0.04-0.13}^{+0.00+0.05}$~MeV and $0.22^{\,+\,0.07\,+\,0.11}_{\,-\,0.06\,-\,0.13}$~MeV, respectively, using events from $b$-hadron decays (the pole on the second Riemann sheet is located at $0.06-i0.13$~MeV relative to the $D^0\bar D^{*0}$ threshold)~\cite{LHCb:2020xds}, and the BESIII reported the mass parameter and imaginary part of $\X$ pole as $\left(3871.63 \pm 0.13_{-0.05}^{+0.06}\right)$~MeV and $(-0.19 \pm 0.08_{-0.19}^{+0.14})$~MeV, respectively, from the processes $e^{+} e^{-} \rightarrow \gamma X(3872)$, with the  $X(3872)$ reconstructed from the $D^0 \bar{D}^0 \pi^0$ and $\pi^{+} \pi^{-} J / \psi$ final states~\cite{BESIII:2023hml}.
One sees that the width from the pole position is much smaller than that from the BW parameterization.

Although the $X(3872)$ has been discovered for more than 20 years, there is still no clear conclusion on what its internal structure is. 
The measurement of the absolute branching fraction ${\cal B}(X(3872)\to \pi^+\pi^- J/\psi)$ brings useful information regarding its complex nature, in particular, leading to insights regarding the isospin symmetry breaking dynamics and possible isospin-1 partners of the $\X$~\cite{Dias:2024zfh}.
In 2019, BaBar first measured the 
branching fraction ${\cal B}(B^+ \to K^+ X(3872))$
with a signal significance of $3\sigma$~\cite{BaBar:2019hzd}, which made the determination of  ${\cal B}(X(3872)\to \pi^+\pi^- J/\psi)$  possible by combining known branching fractions as a product, ${\cal B}(B^+ \to K^+ X(3872))$ ${\cal B}(X(3872)\to \pi^+ \pi^- J/\psi)$, and the results was ${\cal B}(X(3872)\to \pi^+ \pi^- J/\psi)=(4.1\pm 1.3)\%$.
The authors in Ref.~\cite{Li:2019kpj} 
obtained the absolute branching fractions of the six $X(3872)$ decay modes by globally analyzing the measurements from available experiments. In particular, the ${\cal B}(X(3872)\to \pi^+ \pi^- J/\psi)$ was determined to be $(4.1^{+1.9}_{-1.1})\%$, and the dominant decay mode is given by $X(3872) \rightarrow D^{* 0} \bar{D}^0+{\rm c.c.}$\footnote{The charge conjugated component ``c.c.'' will be neglected in the following for simplicity.} with a branching fraction $ \quad\left(52.4_{-14.3}^{+25.3}\right) \%$, implying a strong coupling of the $\X$ to $D\bar D^*$. In addition,
the fraction of the unknown decays of the $X(3872)$ was reported as $(31.9^{+18.1}_{-31.5})\%$. It is desirable to search for more $X(3872)$ decays in the future. 

That the $\X$ mass is so close to the $D^0\bar D^{*0}$ threshold at $(3871.69 \pm 0.07)$~MeV~\cite{ParticleDataGroup:2024cfk} and its strong coupling to the $D^0\bar D^{*0}$ channel led to immediately the proposal that the $\X$ is a hadronic molecule of $D\bar D^{*}$~\cite{Tornqvist:2004qy} predicted by T\"ornqvist in Ref.~\cite{Tornqvist:1993ng}.
Yet, other models, in particular, the compact tetraquark model, for which the $\X$ is a bound state of the $cq$ diquark and $\bar c\bar q$ antidiquark~\cite{Maiani:2004vq} are being discussed (for reviews, we refer to Refs.~\cite{Esposito:2016noz, Guo:2017jvc, Brambilla:2019esw, Chen:2022asf}).

Another physical quantity under active discussion is the ratio of the partial radiative decay widths into $\psi(2S)\gamma$ and $J/\psi \gamma$ final states denoted as 
\begin{eqnarray}
\mathcal{R_{\psi\gamma}}=\frac{\Gamma_{X(3872)\to\psi(2S)\gamma}}{\Gamma_{X(3872)\to J/\psi \gamma}}. 
\end{eqnarray}
A recent measurement of $R_{\psi\gamma}$ from LHCb gives $1.67 \pm 0.21 \pm 0.12 \pm 0.04$, where the first uncertainty is statistical, the second systematic, and the third is
due to the uncertainties on the branching fractions of the $\psi(2S)$ and $J/\psi$ mesons~\cite{LHCb:2024tpv}. This result is consistent with previous measurements from LHCb in 2014~\cite{LHCb:2014jvf}, Belle in 2011~\cite{Belle:2011wdj}, and BaBar in 2008~\cite{BaBar:2008flx}. However, it is quite different from the BESIII result, where an upper limit on the
ratio $R_{\psi\gamma}<0.59$ is set at 90\% confidence level~\cite{BESIII:2020nbj}. This discrepancy needs to be understood in other experiments, especially in Belle II.

The first vector charmonium-like state $Y(4260)$ was observed by BaBar 
via the initial state radiation (ISR) process $e^+e^- \to \gamma_\text{ISR} \pi^+ \pi^- J/\psi$~\cite{BaBar:2005hhc}. Later a few $Y$ states including the $Y(4360)$ and $Y(4660)$ were observed at $B$ factories. To study these  vector charmonium-like states, BESIII collected large data samples above 4 GeV and reported a precise measurement of $e^+e^- \to \pi^+ \pi^- J/\psi$ cross sections from 3.77 to 4.60 GeV~\cite{BESIII:2016bnd}. In the high-statistics analysis by BESIII,  the cross section distribution around 4.26 GeV showed  an asymmetry and resulted in a shift of the peak position of $Y(4260)$ to a lower mass. 
To describe the right shoulder of the $Y(4260)$ line shape, a new BW resonance $Y(4320)$ was introduced, which, however, may also be due to the opening of the $D_1\bar D$ threshold (see, e.g., Ref.~\cite{vonDetten:2024eie}). 
Recently, using data samples with an integrated luminosity of 5.85~fb$^{-1}$, BESIII measured the cross sections for the process $e^+e^- \to K^+ K^- J/\psi$ from 4.61 to 4.95 GeV, where
a new  vector charmonium-like state $Y(4710)$ was observed~\cite{BESIII:2023wqy}. The $Y(4710)$ was also confirmed by BESIII in $e^+e^- \to K_S^0 K_S^0 J/\psi$ with a statistical significance of 4.2$\sigma$~\cite{BESIII:2022kcv}. So far, the $Y(4710)$ is the known vector charmonium-like state with the highest mass. 

The $Z_c(4430)^{\pm}$  reported  by Belle 
in $ B \to  K \pi^{\pm} \psi(2S)$ is
the first good candidate of exotic state with a non-zero electric charge~\cite{Belle:2007hrb}. Although it was not 
confirmed by BaBar in analyzing the same $B$ decays~\cite{BaBar:2008bxw}, LHCb  performed a four-dimensional fit of the decay amplitude for  $B^0 \to K^+ \pi^- \psi(2S)$ 
using $pp$ collision data corresponding to 3 fb$^{-1}$ and found that a highly
significant $Z_c(4430)^{-}$ component is required to describe the data~\cite{LHCb:2014zfx}. Now more charged charmonium-like states including the $Z_c(3900)$~\cite{Belle:2013yex,BESIII:2013ris}, $Z_c(4020)$~\cite{BESIII:2013ouc}, $Z_c(4050)$, $Z_c(4025)$~\cite{Belle:2008qeq}, and even $Z_{cs}(3985)$~\cite{BESIII:2020qkh},
$Z_{cs}(4000)$, $Z_{cs}(4220)$~\cite{LHCb:2021uow} with a strange quark were observed in experiments. 
For a precise determination of the $Z_c(3900)$ pole using modern dispersion theory techniques and discussions about this structure, we refer to Refs.~\cite{Chen:2023def,SCPMA_Guo2}.

Although signals of the $Z_c(3900)$ has been reported in the $J/\psi\pi$ invariant mass distributions, with the $J/\psi\pi^+\pi^-$ invariant mass in the energy region around the $Y(4260)$, by D0 in $p\bar p$ collisions through $b$-flavored hadron decays ~\cite{D0:2018wyb}, no statistically significant signals of the $Y(4260)$ and $Z_c(3900)$ states were observed in the prompt production~\cite{D0:2019zpb}.

\subsection{Hidden charm pentaquark and  double open charm tetraquark}
There is a long history of searches for pentaquark states. The first strong experimental
evidence for a pentaquark state, 
referred to as the $\Theta(1540)^+$, was reported by the LEPS experiment in 
$\gamma n \to n K^+ K^-$~\cite{LEPS:2003wug}.
However, it was not confirmed 
in the same reaction with larger statistics data samples. 

In 2015, LHCb reported the discovery of hidden-charm pentquark candidates in the $\Lambda_b^0 \to K^- p J/\psi$ decay, where the data sample used corresponds to an integrated luminosity of 3 fb$^{-1}$ acquired from 7 and 8~TeV $pp$ collisions~\cite{LHCb:2015yax}. 
A prominent narrow peak was observed in the $p J/\psi$ invariant mass spectrum, and it was named $P_c(4450)^+$. In the amplitude analysis, a second BW resonance, $P_c(4380)^+$, was also introduced to describe the broad bump under the $P_c(4450)$ peak.
The significance of these structures is larger than 9$\sigma$. 
Later LHCb updated this process using the Run 2 data with almost one-order-of-magnitude higher statistics of $\Lambda_b^0$ events. They found that in $p J/\psi$ mass spectrum there are three obvious narrow structures~\cite{LHCb:2019kea}, one for a pentaquark state $P_c(4312)^+$ and two for the $P_c(4440)^+$ and $P_c(4457)^+$, split from the previously observed $P_c(4450)^+$. 

Since the valence structure of $p J/\psi$ is $c\bar{c}uud$, the newly discovered particles must be composed of  at least five (anti)quarks.  
The analysis of the LHCb data in Ref.~\cite{Du:2019pij} suggests that there is also a $1.7\sigma$ significance for a narrow $P_c(4380)$, which is different from the much broader one reported by LHCb in 2015~\cite{LHCb:2015yax}. This analysis is based on the $\Sigma_c^{(*)} \bar{D}^{(*)}$ molecular model, and the existence of such a narrow $P_c(4380)$ is a consequence of heavy quark spin symmetry.
Replacing the $J/\psi$ with a $\eta_c$ meson, LHCb searched for the $P_c(4312)^+$ in $\Lambda_b^0$ decays, but no evidence was observed~\cite{LHCb:2020kkc}.  Since the masses of observed pentaquark states are close to the threshold of a charm baryon and a charm meson, LHCb recently scanned 32 final states including $\Lambda_c^+ \bar{D}$, $\Lambda_c^+ \bar{D}^*$, $\Lambda_c^+  \pi \bar{D}$, 
$\Sigma_c^{(*)}\bar{D}^{(*)}$, $\Lambda_c^+ D$, $\Lambda_c^+ D^*$, $\Lambda_c^+ \pi D$,  $\Sigma_c^{(*)}{D}^{(*)}$, etc., but no significant narrow peak was found for all the modes. 

Two pentaquark candidates with strangeness,
$P_{cs}(4459)^0$ and $P_{cs}(4338)^0$,
were also found in the $J/\psi\Lambda$ system
in $\Xi_b^- \to K^- J/\psi\Lambda$ and 
$B^- \to \bar{p} J/\psi\Lambda$ decays with significances less than and larger than 5$\sigma$, respectively~\cite{LHCb:2020jpq, LHCb:2022ogu}.

Among exotic hadrons, there is a class of hadrons that are particularly interesting, i.e.,  double-charm tetraquarks. 
Such hadrons have been discussed since the 1980s (see Ref.~\cite{Chao:1979tg} for an early treatment with the Born-Oppenheimer approximation).
The first open charm tetraquark, $T_{cc}(3875)^+$ comprised of $cc\bar{u}\bar{d}$, was discovered in 
the $D^0D^0 \pi^+$ invariant mass spectrum by LHCb in 2021~\cite{LHCb:2021vvq,LHCb:2021auc}.
Its mass is just below the $D^0 D^{*+}$ mass threshold, while the width is only a few tens of keV.

\section{Production in nuclear collisions}
\label{sec:exp_HIC}
\subsection{Existing and future experiments}
\label{sec:exp_HIC1}

High-energy collisions of heavy ions are a powerful method for generating extremely hot and dense nuclear matter, often referred to as quark-gluon plasma (QGP)~\cite{Shuryak:1980tp,Chen:2024zwk, Shou:2024uga}, which exhibits an energy density comparable to that of the universe just a few microseconds after the Big Bang, with roughly equal numbers of quarks and antiquarks. The extreme energy density of the QGP phase leads to the creation of many strange-antistrange quark pairs from the quantum vacuum. As the QGP cools, it transites into a hadron gas, resulting in the formation of various baryons, mesons, and their antiparticles. 
Therefore, these collisions offer a unique opportunity to explore exotic particles, such as antimatter, hypernuclei~\cite{STAR:2010gyg,STAR:2011eej,STAR:2015kha,ALICE:2015oer,ALICE:2017jmf,STAR:2019wjm,STAR:2023fbc,Chen:2018tnh}, and exotic hadrons~\cite{ExHIC:2010gcb}, thereby uncovering important fundamental interactions.

\begin{figure}[tbh]
\centering
\includegraphics[width=\linewidth]{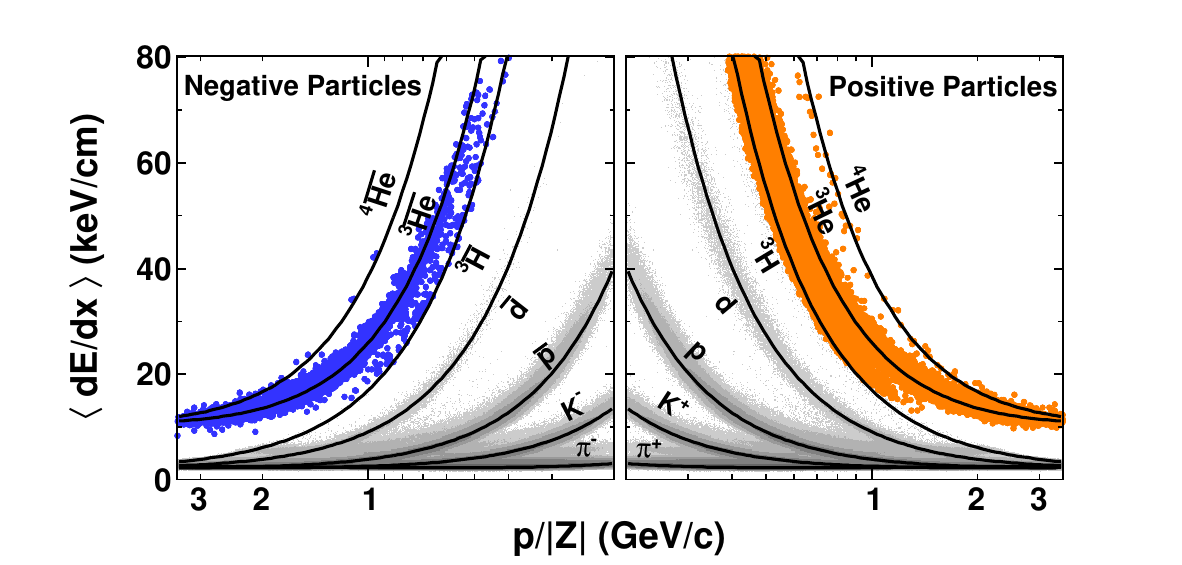}
\caption{TPC mean energy loss per unit track length, $\langle dE/dx \rangle$ as a function of rigidity, $p/|Z|$ with $p$ and $Z$ being momentum and charge, respectively. From Ref.~\cite{STAR:2011eej}.}
\label{fig:dedx}
\end{figure}

While nuclei are abundant across the universe, antinuclei heavier than the antiproton have only been observed as products of relativistic heavy-ion collisions at facilities like the BNL Relativistic Heavy-Ion Collider (RHIC) and the CERN Large Hadron Collider (LHC). In these experiments, the time projection chamber (TPC) positioned within a solenoidal magnetic field plays a crucial role in identifying the particles of interest by measuring the mean energy loss per unit track length $\langle dE/dx\rangle$, as illustrated in Fig.~\ref{fig:dedx}. In the left panel, STAR reported the observation of \(^4\overline{\text{He}}\), the heaviest stable antinucleus detected to date~\cite{STAR:2011eej}. A total of 18 counts were detected in Au+ Au collisions at center-of-mass (c.m.) energies of 200 GeV and 62 GeV per nucleon–nucleon pair. This discovery comes a century after Rutherford observed the alpha particle. The yield also aligns with predictions from both thermodynamic and coalescent nucleosynthesis models~\cite{STAR:2011eej}.

Hypernuclei are bound states of nucleons and hyperons, and they offer valuable insights into the hyperon-nucleon interaction. While nucleons (proton and neutron) are composed solely of up and down quarks, hyperons are light-flavor baryons containing at least one strange quark. A hypernucleus is defined as a nucleus that contains at least one hyperon in addition to nucleons. Despite being bound within hypernuclei, all hyperons are inherently unstable {since they can decay via weak interaction}. The simplest bound hypernucleus is the hypertriton (\(^3_\Lambda \text{H}\))~\cite{Chen:2023mel}, which comprises a \(\Lambda\) hyperon, a proton, and a neutron. The first hypernucleus was observed in 1952 using a nuclear emulsion cosmic ray detector~\cite{Danysz:1953}. In contrast, the first observation of antihypertritons---composed of an antiproton, an antineutron, and an $\bar{\Lambda}$---was reported by STAR in Au+Au collisions at $\sqrt{s_{NN}}=$ 200 GeV in 2010~\cite{STAR:2010gyg}, through the decay channel \( ^3_{\bar{\Lambda}}\overline{\text{H}} \rightarrow {^3\overline{\text{He}}} + \pi{^+} \). The measured yields of \(^3_{\bar{\Lambda}}\overline{\text{H}}\) and 
\(^3\overline{\text{He}}\) are comparable, indicating a balance in the populations of up, down, and strange quarks and antiquarks in both coordinate and momentum space~\cite{STAR:2010gyg,STAR:2023fbc} as a consequence of the approximate SU(3) flavor symmetry among the up, down and strange quarks of QCD.

\begin{figure}[tb]
    \centering
    \includegraphics[width=0.61\linewidth]{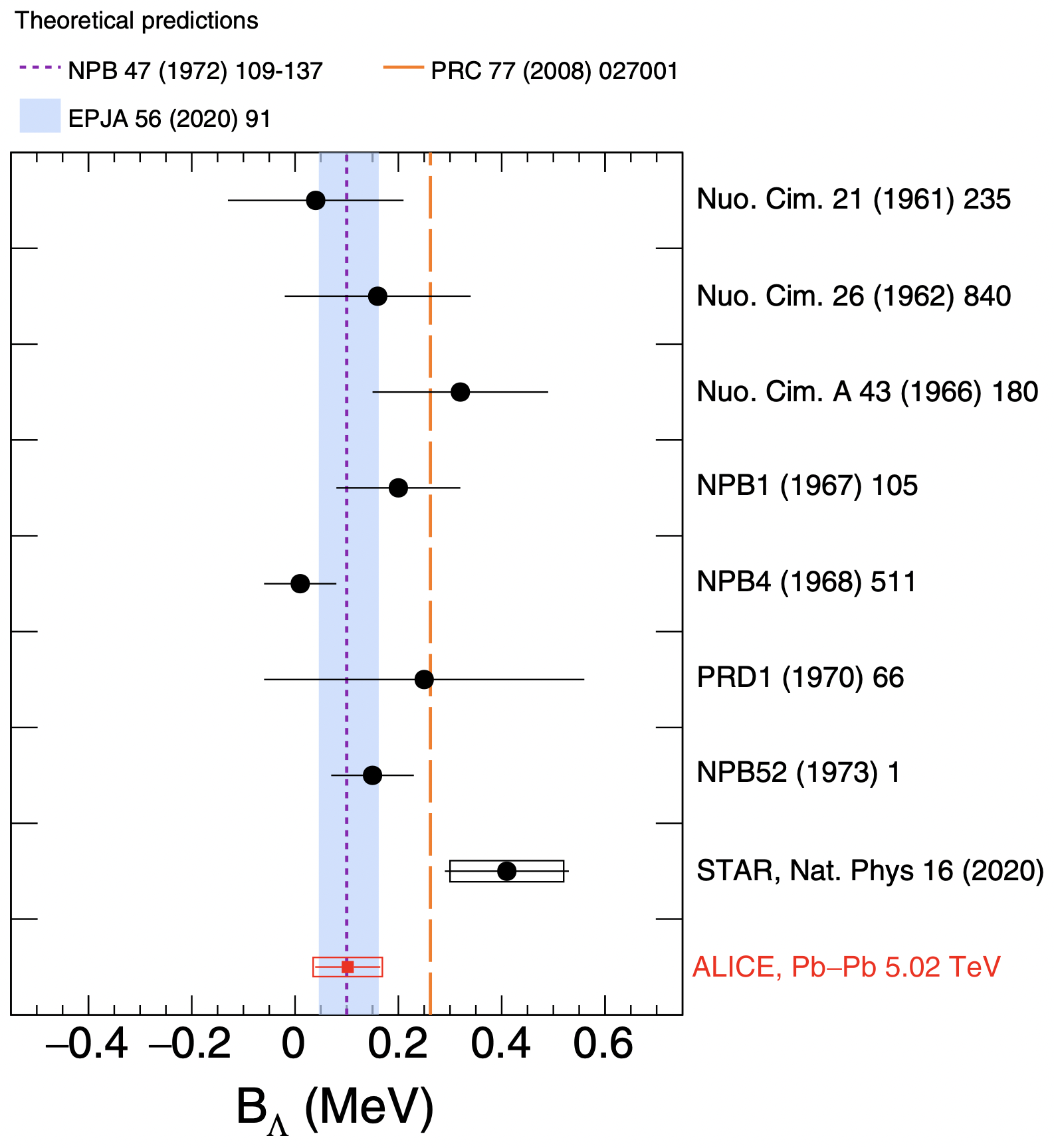}
    \includegraphics[width=0.38\linewidth]{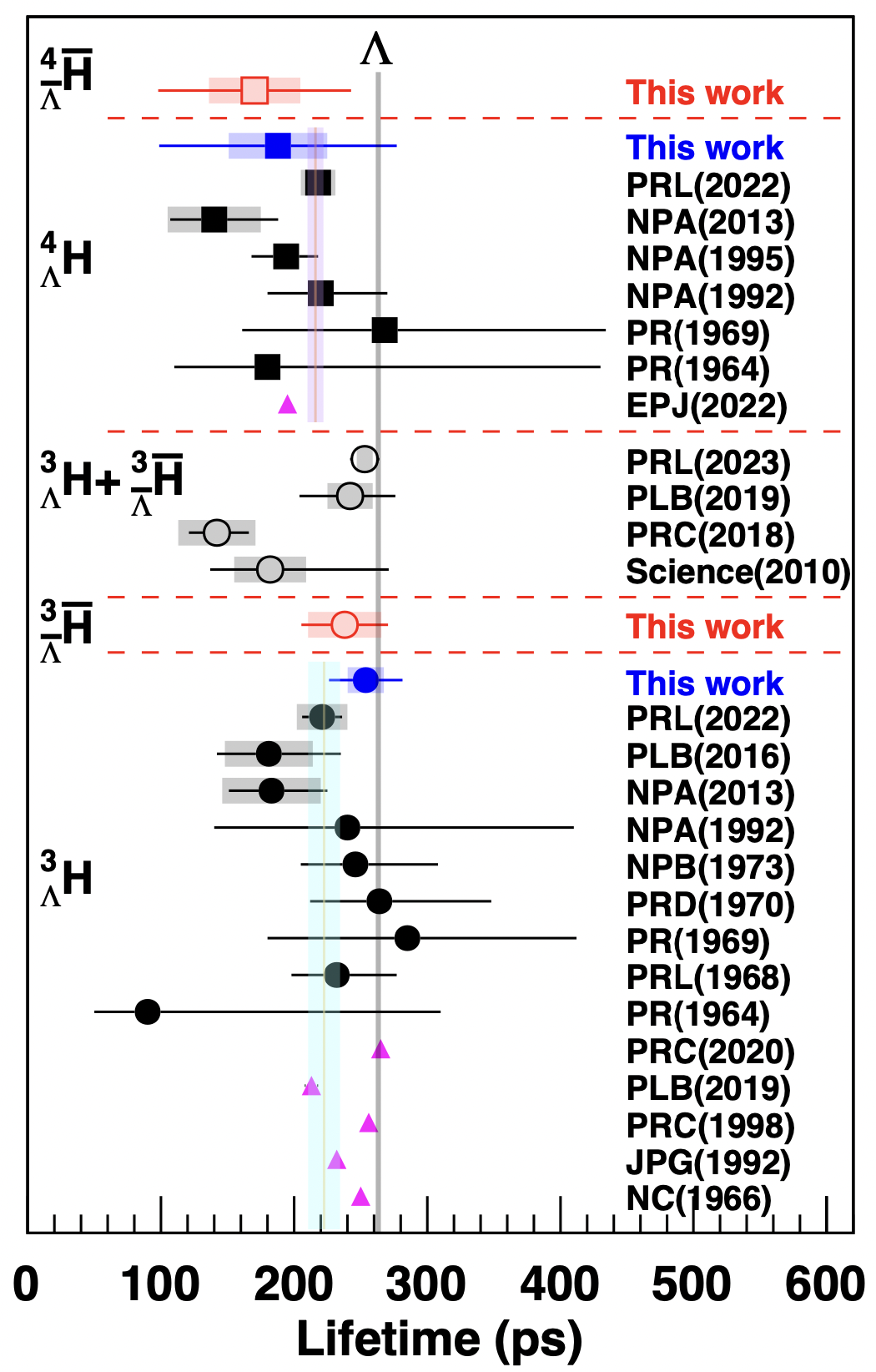}
    \caption{The \(\Lambda\) separation energy \(B_{\Lambda}\) of hypertriton (left) and the lifetime of various hypernuclei (right) measured in different experiments. From Refs.~\cite{ALICE:2022sco} and \cite{STAR:2023fbc}.}
    \label{fig:ALICEHypertriton}
\end{figure}

The most precise measurements of the \(^3_\Lambda\text{H}\) lifetime (\(\tau\)) and the \(\Lambda\) separation energy (\(B_\Lambda\)) were obtained in the Pb-Pb collisions at \(\sqrt{s_{\rm NN}} = 5.02\) TeV at ALICE~\cite{ALICE:2022sco} (c.f. Fig.~\ref{fig:ALICEHypertriton}). These measurements are consistent with predictions from effective field theories and support the characterization of \(^3_\Lambda\text{H}\) as a weakly bound system~\cite{Chen:2023mel}.
Although some discrepancies, referred to as the "hypertriton puzzle," have been noted in the literature concerning the lifetime and \(B_\Lambda\), a global average of all available measurements reveals no significant global tension, with a 23\% probability for the lifetime and 57\% for \(B_\Lambda\), as determined by a Pearson test~\cite{ALICE:2022sco}. The upcoming Run 3 of the LHC is expected to provide these measurements with an unprecedented precision.

Recently, STAR reported the observation of the antimatter hypernucleus \(^4_{\bar{\Lambda}}\overline{\text{H}}\), which consists of an \(\bar{\Lambda}\), an antiproton, and two antineutrons~\cite{STAR:2023fbc}. This discovery was achieved through its two-body decay. A total of 15.6 candidate \(^4_{\bar{\Lambda}}\overline{\text{H}}\) antimatter hypernuclei were identified, with an estimated background count of 6.4~\cite{STAR:2023fbc}. This was the heaviest unstable antinucleus observed so far. The lifetimes of the antihypernuclei \(^3_{\bar{\Lambda}}\overline{\text{H}}\) and \(^4_{\bar{\Lambda}}\overline{\text{H}}\) were measured and compared with those of their corresponding hypernuclei, providing a test of the symmetry between matter and antimatter. The lifetimes of the (anti)hypernuclei \(^3_{\bar{\Lambda}}\overline{\text{H}}\), \(^4_{\bar{\Lambda}}\overline{\text{H}}\), \(^3_{\Lambda}\text{H}\), and \(^4_{\Lambda}\text{H}\) are depicted in Fig.~\ref{fig:ALICEHypertriton}. The differences in lifetime between hypernuclei and their corresponding antihypernuclei are \(\tau(^3_{\bar{\Lambda}}\overline{\text{H}}) - \tau(^3_{\Lambda}\text{H}) = 16 \pm 43 (\text{stat.}) \pm 20 (\text{sys.})\) ps and \(\tau(^4_{\bar{\Lambda}}\overline{\text{H}}) - \tau(^4_{\Lambda}\text{H}) = 18 \pm 115 (\text{stat.}) \pm 46 (\text{sys.})\) ps. Both differences are consistent with zero within uncertainties, indicating no significant disparity between the properties of matter and antimatter particles, thus providing a new test of CPT symmetry. 

\begin{figure}[tb]
    \includegraphics[width=\linewidth]{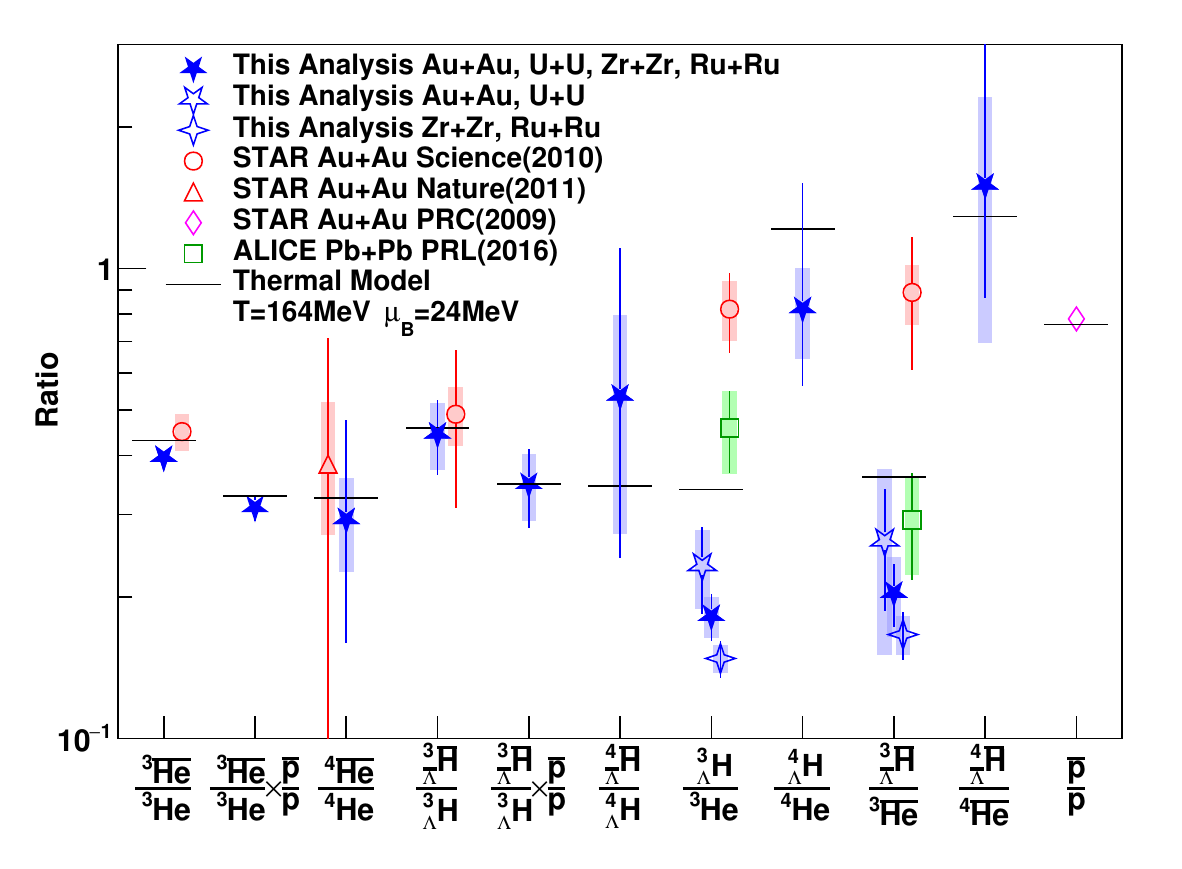}
    \caption{Production yield ratios among the various (anti)nuclei and (anti) hypernuclei with the same number of (anti)baryons measured in heavy-ion collisions. From Ref.~\cite{STAR:2023fbc}.}
    \label{fig:STAR4H}
\end{figure}

Various production yield ratios among (anti)hypernuclei (including hypernuclei and/or antihypernuclei) and (anti)nuclei (including nuclei and/or antinuclei) were also assessed and compared with theoretical models, offering insights into their production mechanisms. As shown in Fig.~\ref{fig:STAR4H}, we see that the antimatter-over-matter particle-yield ratios are measured to be below unity because the colliding heavy ions carry positive baryon numbers, and consequently the collision system has positive baryon chemical potential. The data are consistent with most existing measurements within their uncertainties and are consistent with the expectation of the coalescence picture of (anti)nucleus and (anti)hypernucleus production and the statistical thermal
model~\cite{STAR:2023fbc}. 
The relations among the production yield ratios \(^4\overline{\text{He}}/^4\text{He} \approx {^3\overline{\text{He}}}/^3\text{He} \times \bar{p}/p\), 
\(^4_{\bar\Lambda}\overline{\text{H}}/^4_{\Lambda}\text{H} \approx {^3_{\bar\Lambda}\overline{\text{H}}}/^3_{\Lambda}\text{H} \times \bar{p}/p\), \(^4_{\Lambda}\text{H}/^4\text{He} \approx 4 \times {^3_{\Lambda}\text{H}}/^3\text{He}\), and \(^4_{\bar{\Lambda}}\overline{\text{H}}/^4\overline{\text{He}} \approx 4 \times {^3_{\bar{\Lambda}}\overline{\text{H}}}/^3\overline{\text{He}}\) are consistent with the coalescence model of (anti)nucleus and (anti)hypernucleus production. 
Here the factor of 4 in the last two relations arises because both spin-0 and spin-1 states of \(^4_\Lambda\text{H}\) possess sufficiently large binding energies, leading to no energetically allowed strong decay channels. 
Consequently, the spin-1 state, with a spin degeneracy of 3, decays electromagnetically to the spin-0 ground state, enhancing the total production yields of \(^4_\Lambda\text{H}\) and \(^4_{\bar\Lambda}\overline{\text{H}}\) by a factor of 4 compared to those of \(^4\text{He}\) and \(^4\overline{\text{He}}\), which have only spin-0 states.
Considering this spin-degeneracy effect, the statistical-thermal model predictions align well with experimental measurements, although the \(^4_{\Lambda}\text{H}/^4\text{He}\) ratio is slightly lower than  the predicted value. This discrepancy, though not statistically significant, might be attributed to the very small binding energy of \(^3_{\Lambda}\text{H}\), suggesting that the spatial extent of its wave function is comparable to the entire collision system~\cite{Liu:2024ygk}.

High energy $pp$ and heavy-ion collisions is also an excellent laboratory to explore multistrange dibaryons~\cite{Shah:2015oha,Shao:2020sqr, Zhang:2020dma, Zhang:2021vsf, MaYG}. In the search for the possible $\Lambda\Lambda$ bound state, known as the H-dibaryon, femtoscopic correlations of $\Lambda\Lambda$ pairs are studied in $pp$, Au+Au and $p$Pb collisions~\cite{STAR:2014dcy,ALICE:2019eol}. By comparing measured data with model calculations, the scattering parameter space, characterized by the inverse scattering length and effective range, is constrained. The data reveal a shallow attractive interaction, consistent with findings from hypernuclei studies and lattice computations.

In addition to antimatter and hypernuclei, heavy-ion collisions also serve as a laboratory to study the hidden-charm $XYZ$ particles. The first evidence of \( X(3872) \) production in relativistic heavy-ion collisions has been reported~\cite{CMS:2021znk}. The \( X(3872) \) production was investigated in Pb-Pb collisions at \(\sqrt{s_{NN}} = 5.02\) TeV, utilizing the decay chain \( X(3872) \rightarrow J/\psi \pi^+ \pi^- \rightarrow \mu^+ \mu^- \pi^+ \pi^- \). The significance of the inclusive \( X(3872) \) signal was measured at 4.2 standard deviations~\cite{CMS:2021znk}. The prompt \( X(3872) \) to \(\psi(2S)\) yield ratio was found to be \( \rho_\text{Pb-Pb} = 1.08^{+0.49}_{-0.52} \), significantly higher compared to typical values of 0.1 observed in $pp$ collisions~\cite{CMS:2021znk}.
For more discussions, see Section~\ref{sec:HIC_th}.

\begin{figure}[tb]
    \centering
    \includegraphics[width=0.43\linewidth]{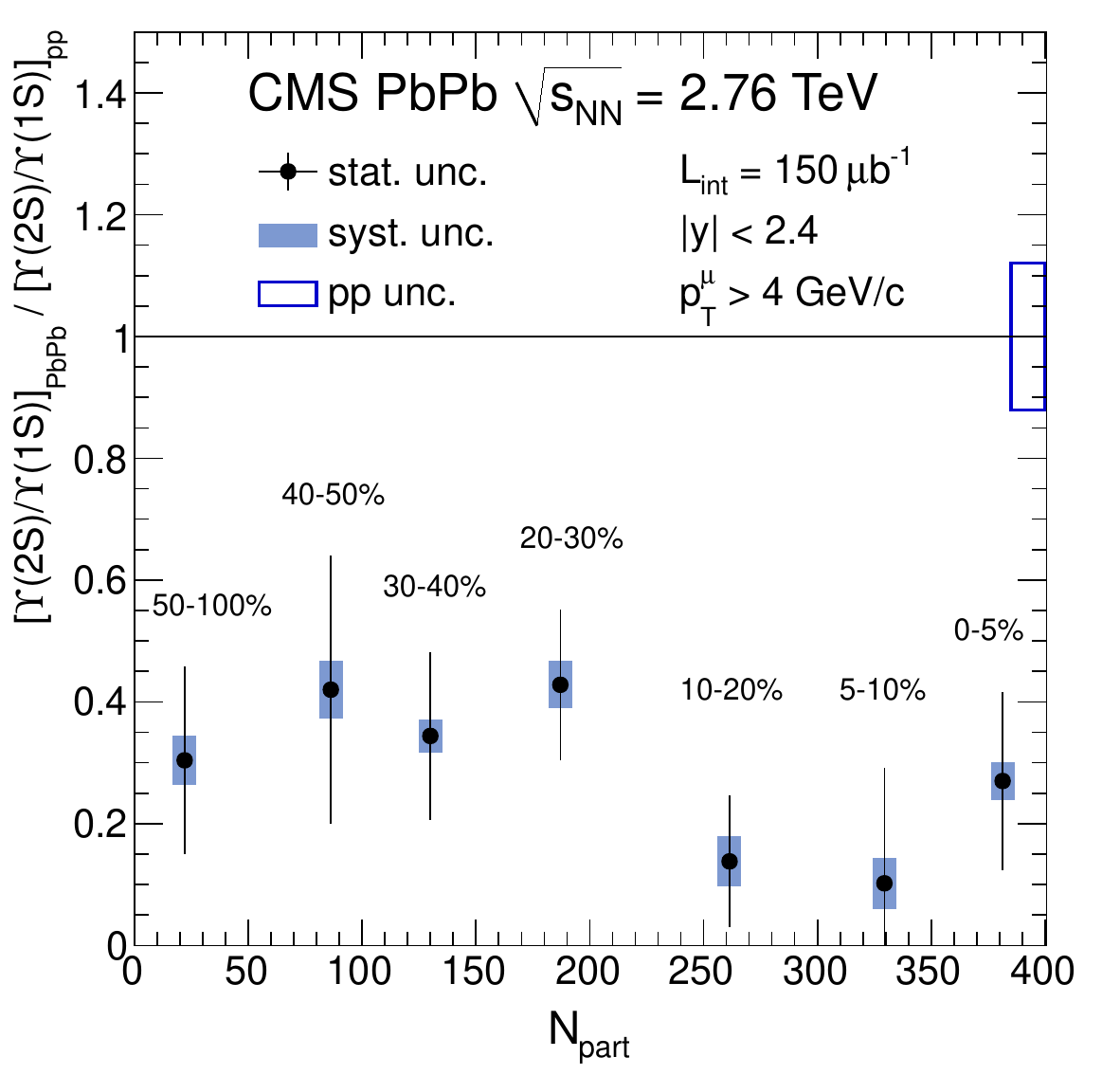}
    \includegraphics[width=0.55\linewidth]{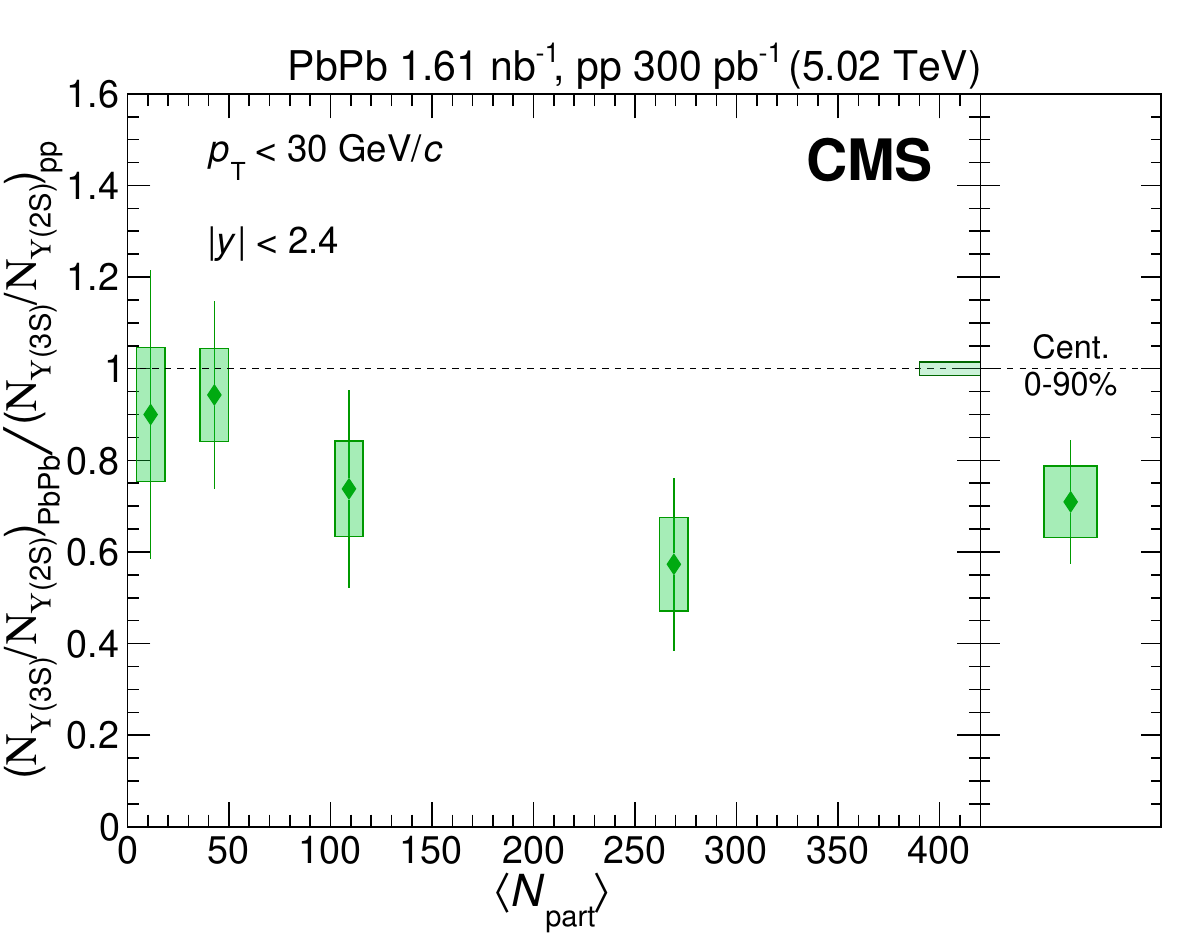}
    \includegraphics[width=0.43\linewidth]{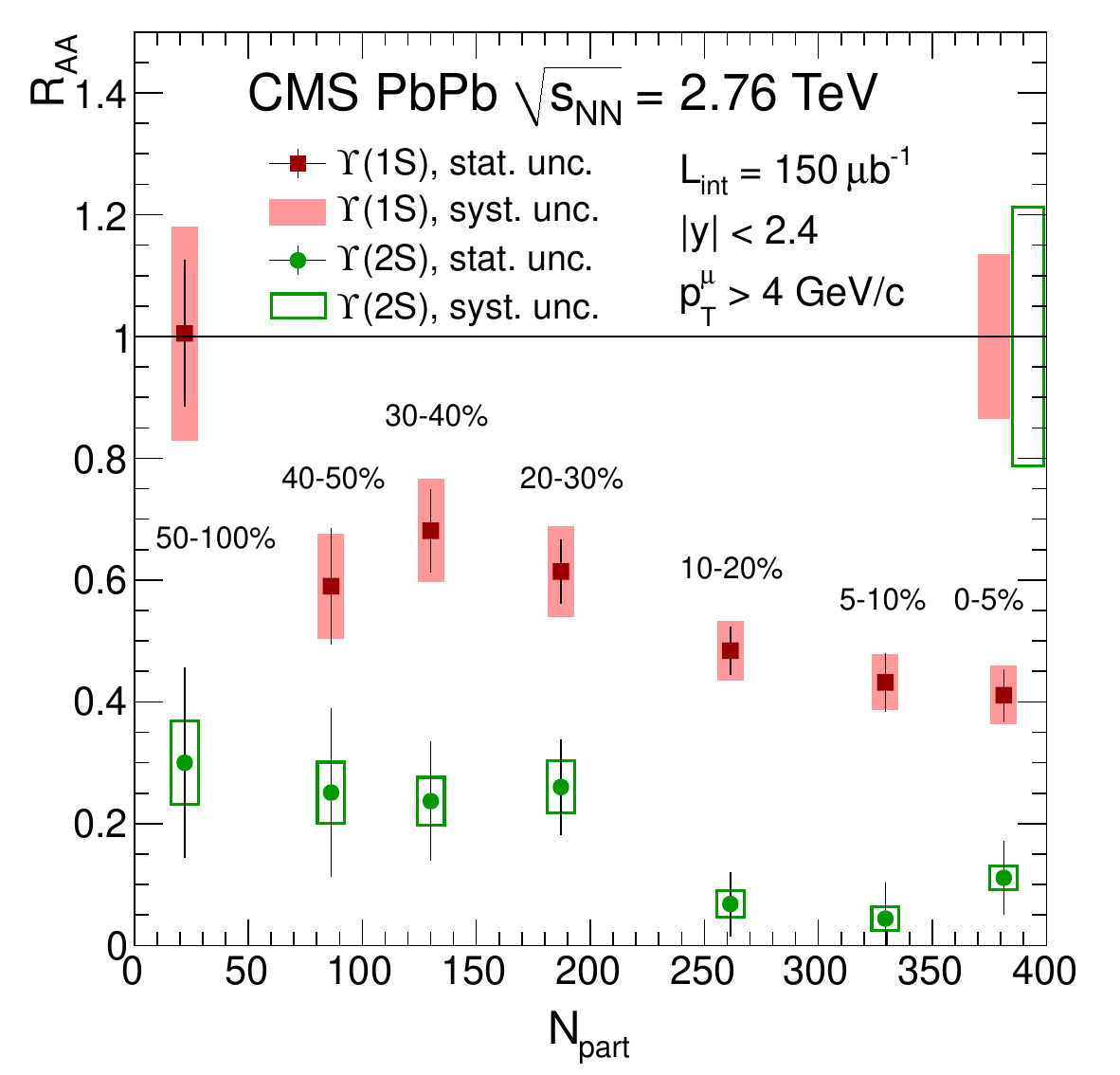}
    \includegraphics[width=0.55\linewidth]{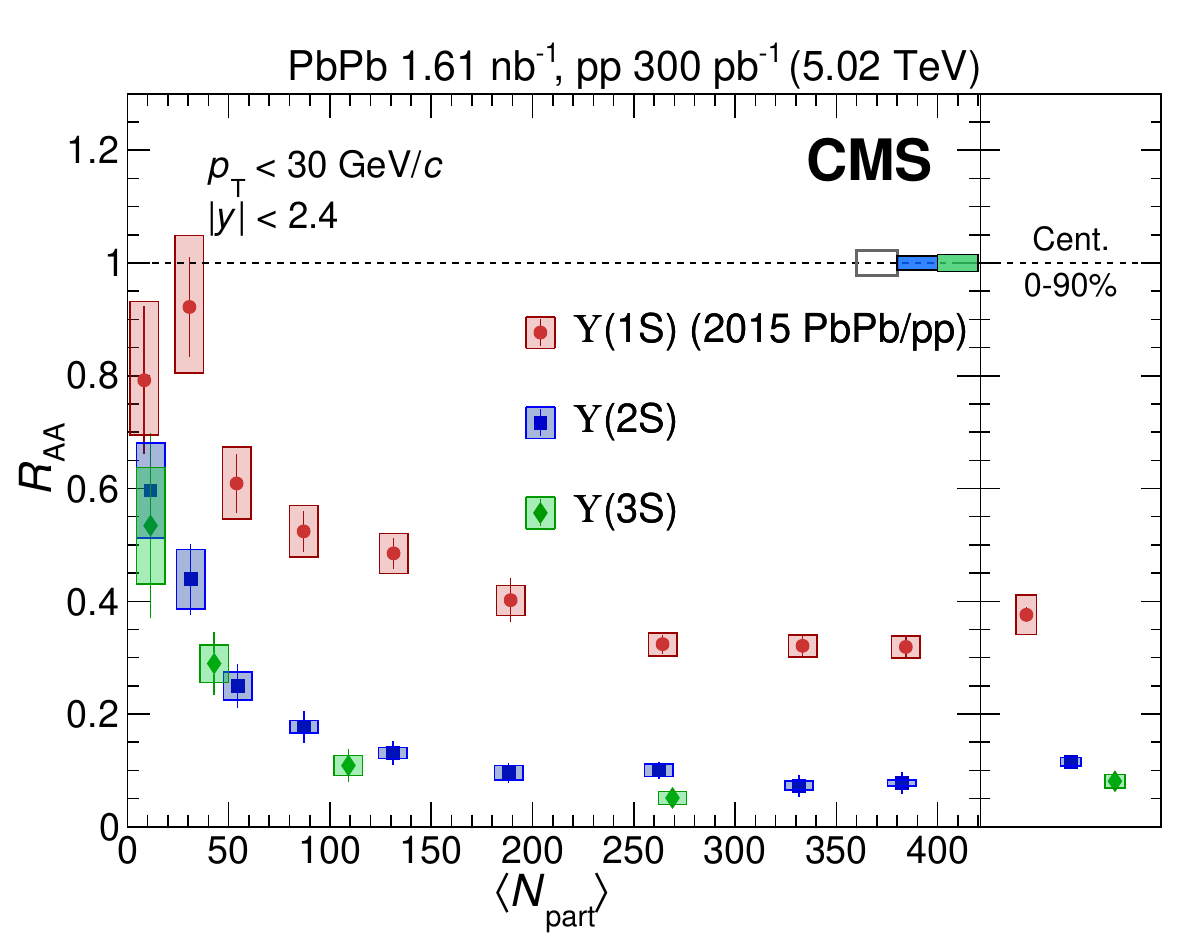}
    \caption{The double ratios among different \(\Upsilon(nS)\) states and nuclear modification factor $R_{AA}$ as functions of \(\langle N_{\text{part}} \rangle\). From Refs.~\cite{CMS:2012gvv,CMS:2023lfu}.}
    \label{fig:RAAv2}
\end{figure}
    
Another common observable regarding the interaction of quarkonium states with medium created in heavy-ion collisions is the nuclear modification factor \(R_{AA}\), defined as the ratio of the yield in central heavy-ion collisions to that in $pp$ collisions, normalized by the number of binary collisions in the reaction. In Refs.~\cite{CMS:2016rpc,CMS:2018zza,CMS:2012gvv,CMS:2023lfu}, the suppression of the production yields of the \(\psi(nS)\) and \(\Upsilon(nS)\) mesons in Pb-Pb collisions relative to those in $pp$ collisions have been studied using data from the CMS experiment at the LHC, which are summarized in Fig.~\ref{fig:RAAv2}.
For the \(\psi(nS)\) states, integrated over collision centralities, the nuclear modification factors \(R_{AA}\) are $0.56 \pm 0.08$ (stat.) $\pm 0.07$ (syst.) for \(J/\psi\), $0.12 ± 0.04$ (stat.) $\pm 0.02$ (syst.) for \(\psi(2S)\), and less than 0.10 (at 95\% confidence level) for \(\psi(3S)\), revealing a sequential suppression~\cite{CMS:2016rpc,CMS:2018zza}. The \(\Upsilon(2S)\) and \(\Upsilon(3S)\) mesons were also studied~\cite{CMS:2016rpc,CMS:2018zza}, with the \(\Upsilon(3S)\) observed for the first time in Pb-Pb collisions with a significance above 5 standard deviations. The suppression of \(\Upsilon\) yields increases from peripheral to central collisions and is more pronounced for \(\Upsilon(3S)\) compared to \(\Upsilon(2S)\) (see Fig.~\ref{fig:RAAv2}), extending the pattern of sequential suppression previously noted for other quarkonium states.

\begin{figure}[tb]
    \centering
    \includegraphics[width=0.53\linewidth]{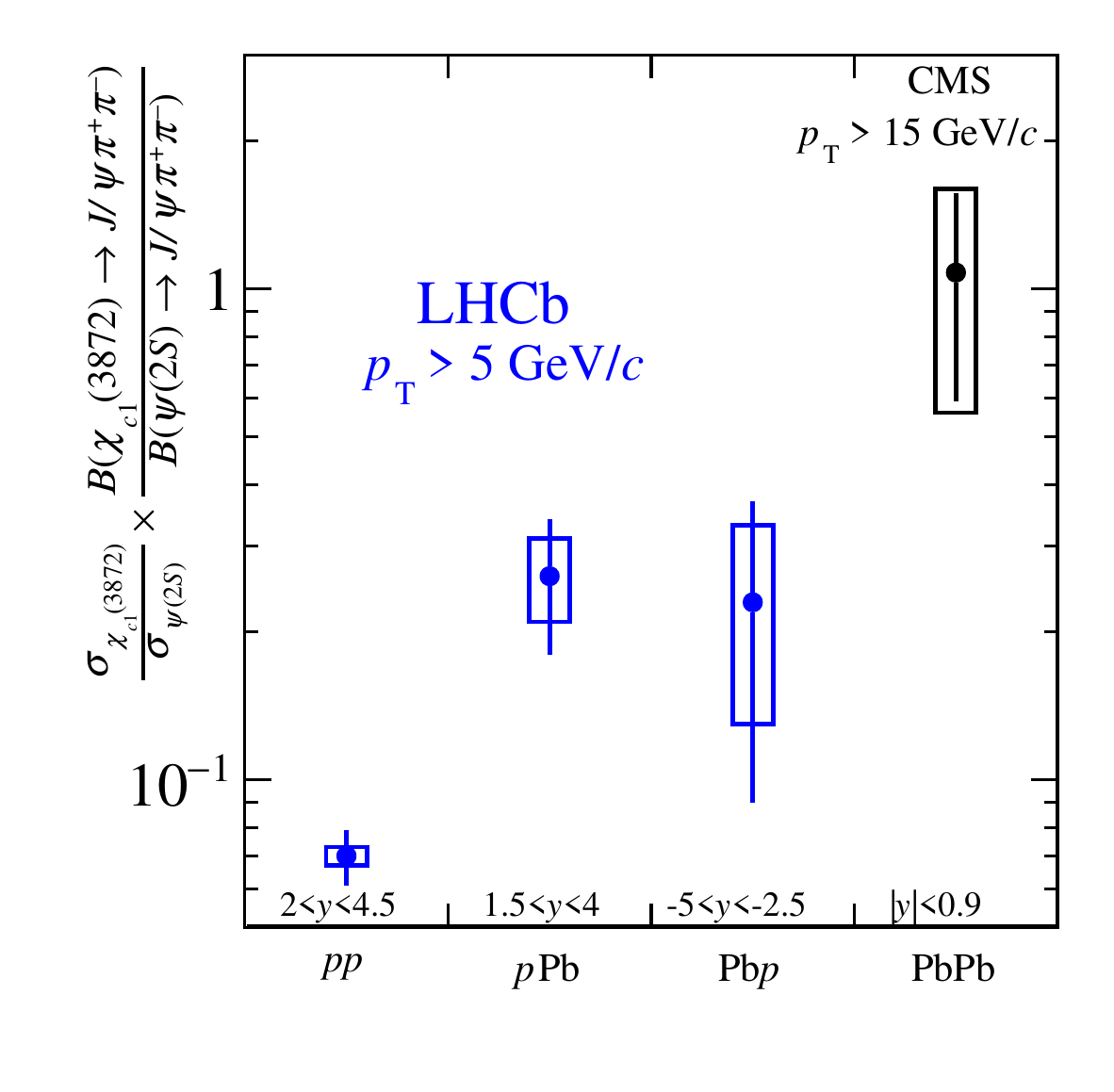}
    \includegraphics[width=0.46\linewidth]{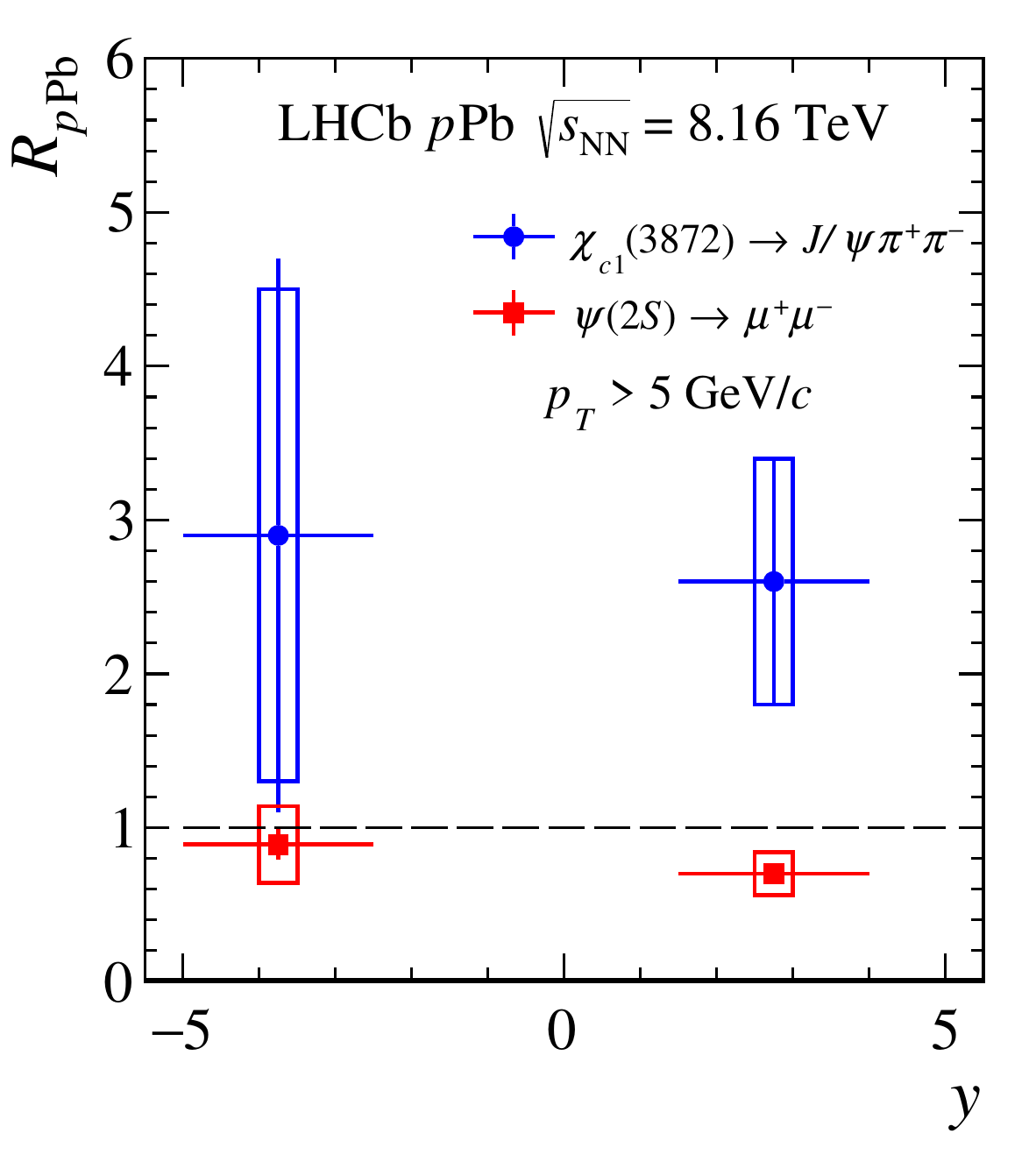}
    \caption{Left: The ratio of the production cross sections for  \(\X\) to \(\psi(2S)\) in the \(J/\psi \pi^+ \pi^-\) decay channel, measured in various collision systems. Right: The nuclear modification factor \( R_{p\text{Pb}} \) for the \(\X\) and \(\psi(2S)\) hadrons. From Ref.~\cite{LHCb:2024bpb}.}
    \label{fig:R_Xpsi}
\end{figure}

In Ref.~\cite{LHCb:2024bpb}, LHCb has measured the production of the exotic hadron \(\X\) in $p$Pb collisions at \(\sqrt{s_{NN}} = 8.16\) TeV. Comparing this with the charmonium state \(\psi(2S)\) reveals that the dynamics of \(\X\) in the nuclear medium differ from those of conventional hadrons. Additionally, when compared to proton-proton collision data, it appears that the presence of the nucleus may affect the production rates of \(\X\) (see Fig.~\ref{fig:R_Xpsi}). This marks the first determination of the nuclear modification factor for an exotic hadron.

\begin{figure}[tb]
    \includegraphics[width=0.9\linewidth]{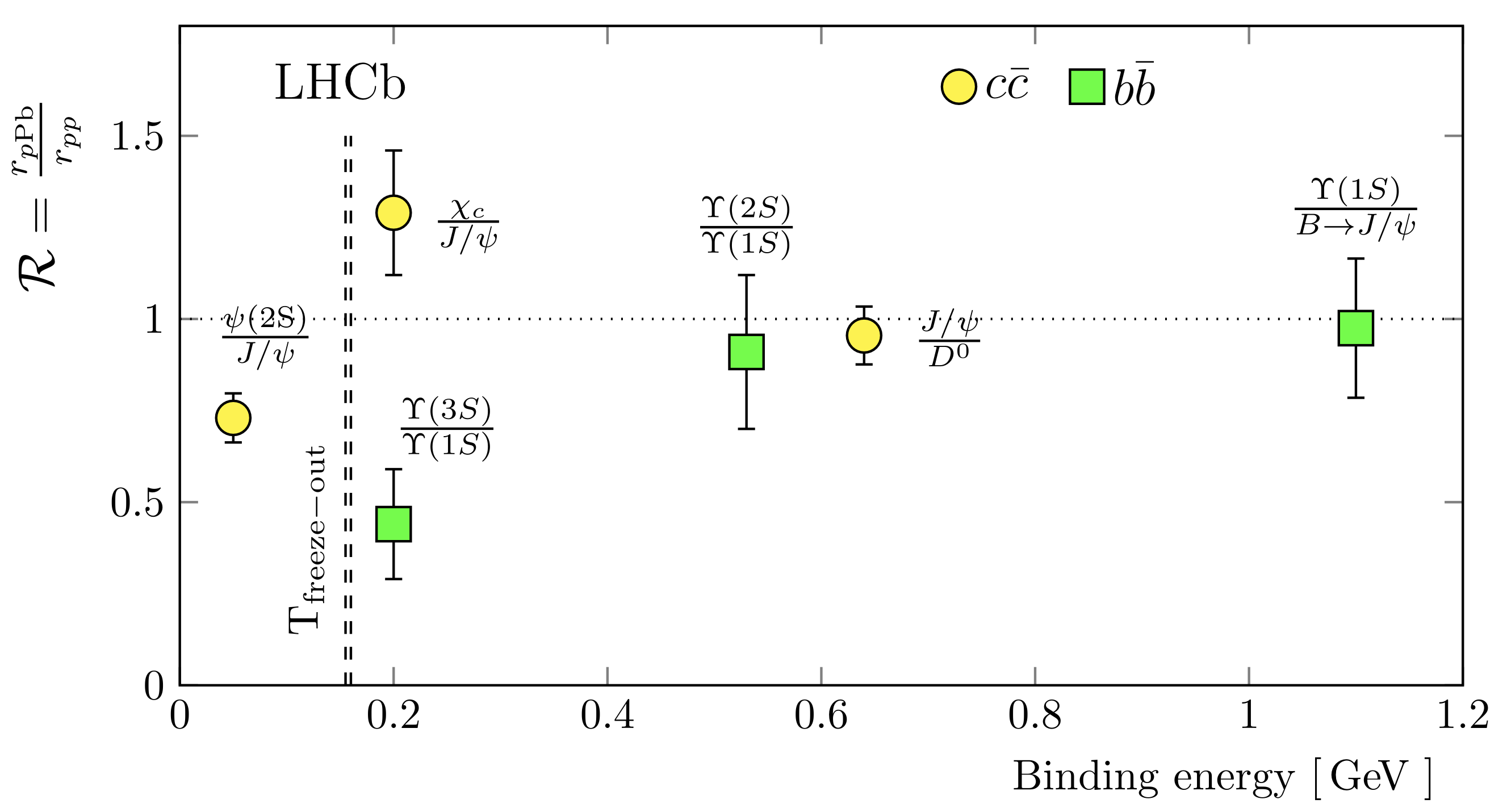}
    \caption{The double ratio of the quarkonium state yield relative to its ground state or open heavy quark meson in \(p\)Pb compared to \(pp\) collisions as a function of the binding energy of the quarkonium state. From Ref.~\cite{LHCb:2023apa}.}
    \label{fig:heavyIon2}
\end{figure}    

LHCb has also measured the feed-down fraction of \(\chi_{c1}\) and \(\chi_{c2}\) decays contributing to the prompt \(J/\psi\) yield in $p$Pb collisions at \(\sqrt{s_{NN}} = 8.16\) TeV. The results, presented as a function of \(J/\psi\) transverse momentum \(p_{T,J/\psi}\) in the range 1 $<$ \(p_{T,J/\psi}\) $<$ 20 GeV, show that the fraction at forward rapidity is consistent with the measurement in $pp$ collisions at \(\sqrt{s} = 7\)~TeV~\cite{LHCb:2012af} (see Fig.~\ref{fig:heavyIon2}). However, the fraction at backward rapidity is 2.4$\sigma$ higher than that at at forward rapidity for \(1 < p_{T,J/\psi} < 3\)~GeV. This increase at low \(p_{T,J/\psi}\) in the backward region aligns with the observed suppression of \(\psi(2S)\) contributions to the prompt \(J/\psi\) yield. Additionally, the study places an upper limit of 180 MeV on the free energy available in these $p$Pb collisions that could dissociate or inhibit the formation of charmonium states, indicating no significant in-medium dissociation of the \(\chi_{cJ}\, (J=1,2)\) states~\cite{LHCb:2023apa}.

\begin{figure}[tb]
\centering
\includegraphics[width=0.49\linewidth]{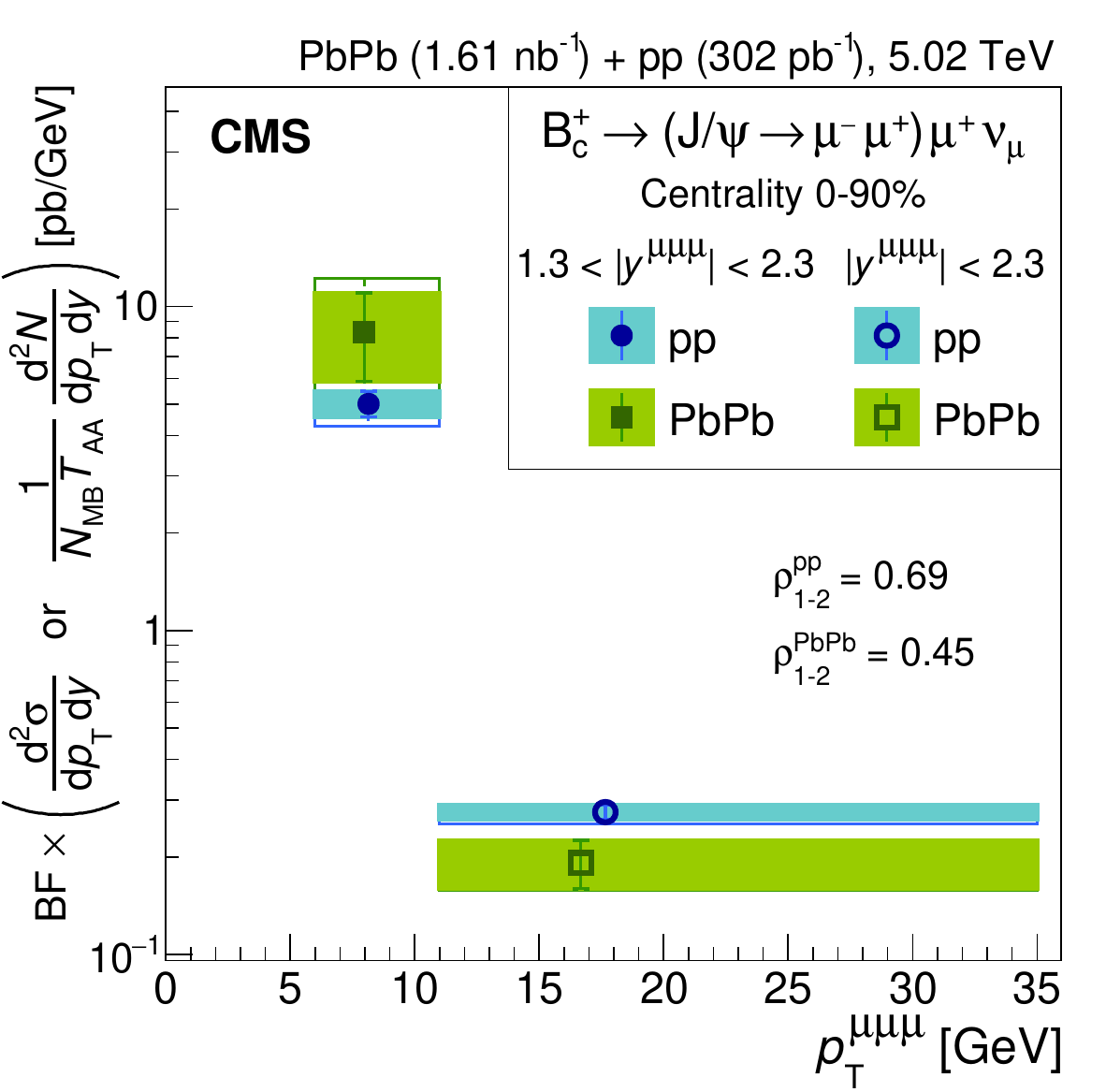}
\includegraphics[width=0.49\linewidth]{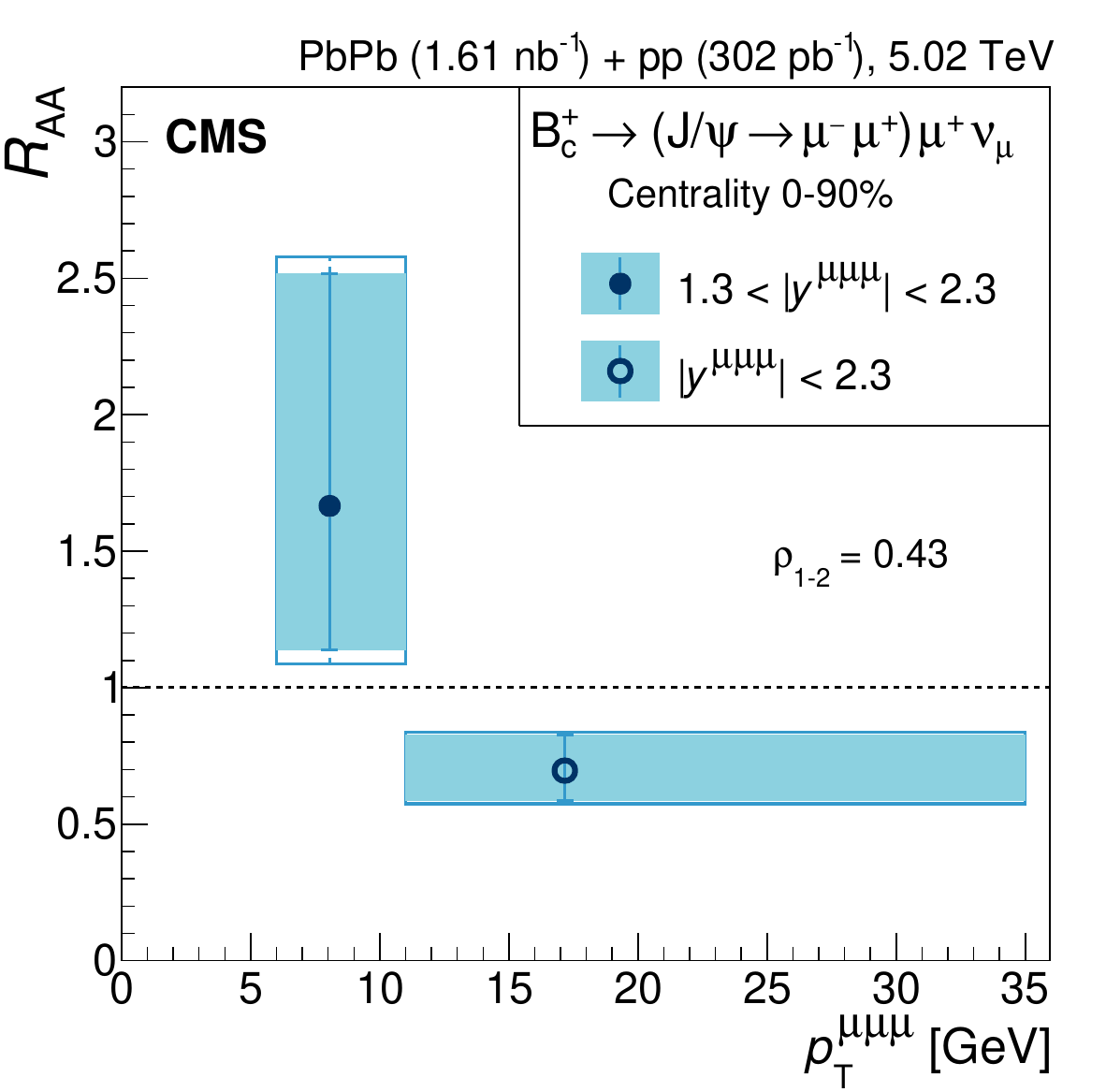}
\caption{$B^+_c$ meson production in $pp$ and Pb-Pb collisions at LHC energies. From Ref.~\cite{CMS:2022sxl}.}
\label{fig:Bc}
\end{figure}

LHC detectors also observed the \(B^+_c\) meson in heavy-ion collisions~\cite{CMS:2022sxl}. CMS examines the production of the \(B^+_c\) meson in Pb-Pb and $pp$ collisions at \(\sqrt{s_{NN}} = 5.02\) TeV, through its decay channel \(B^+_c \to (J/\psi \to \mu^+ \mu^-) \mu^+ \nu_\mu\)~\cite{CMS:2022sxl}. 
The left panel of Fig.~\ref{fig:Bc} presents the measured \(B^+_c\) cross section.
The two bins of the trimuon $p_T$ correspond to different rapidity ranges (see Ref.~\cite{CMS:2022sxl} for more details.). The ratio between the low $p_T$ and high $p_T$ regions is $18.2^{+1.3}_{-2.1}$ in $pp$ data and 24.1 in simulation, suggesting that the simulation overestimates the spectrum steepness~\cite{CMS:2022sxl}. 
The right panel of Fig.~\ref{fig:Bc} shows the nuclear modification factor for the \(B^+_c\) meson, obtained by comparing the production cross sections in Pb-Pb to $pp$ collisions, measured across different transverse momentum and collision centrality bins. 
The \(B^+_c\) meson shows less suppression than quarkonia and most open heavy-flavor mesons, indicating that the hot and dense nuclear matter created in heavy-ion collisions may enhance its production~\cite{CMS:2022sxl}. 
This finding provides a valuable new probe into the complex dynamics of suppression and enhancement mechanisms of heavy quarkonia in hot dense matter. This is another connection on the study of the structure of heavy quarkonium(-like) states and the heavy-ion collisions.

\subsection{Production in \texorpdfstring{$\bm{pp/p\bar p}$}{pp/p pbar} collisions}  
\label{sec:prod_pp}

\subsubsection{Exotic hadrons with heavy quarks}
\label{sec:Xin_pp}

Exotic hadrons can be produced in $pp/p\bar p$ collisions mainly in two ways, i.e., in weak decays of $b$-flavored hadrons and in prompt processes. 
The former, through the $b\to c\bar c s$ process at the quark level, is the prominent source of exotic hadrons, with examples like the $\X$, the $Z_c(4430)$ and the several $P_c$ states discussed in Section~\ref{sec:multiquarkexp}. 
The latter produces exotic hadrons directly through strong interactions, which is the main focus of this section.

Shortly after the discovery of the $\X$ by the Belle Collaboration in $B$ decays~\cite{Belle:2003nnu}, the CDF and D0 Collaborations reported the observation of the $\X$ in $p\bar p$ collisions in semi-inclusive processes~\cite{CDF:2003cab, D0:2004zmu}.
Later on, the semi-inclusive productions of the $\X$ in $pp$ collisions were observed by the CMS and LHCb Collaborations~\cite{LHCb:2011zzp, CMS:2013fpt, LHCb:2020sey}.
All these observations were made in the $J/\psi\pi^+\pi^-$ final state.

Using the CDF measurements of the yields of the $\X$~\cite{CDF:2006ocq} and $\psi(2S)$~\cite{CDF:2009kwm}, the prompt production rate of the $\X$ in $p\bar p$ collisions at 1.96~TeV was estimated to be~\cite{Bignamini:2009sk, Artoisenet:2009wk}
\begin{align}
    \sigma[p\bar p\to X(3872)+\text{all}] &\operatorname{Br}\left[X \rightarrow J / \psi \pi^{+} \pi^{-}\right] \notag\\
    &\approx (3.1 \pm 0.7)~\mathrm{nb},
\end{align}
where the undetected particles produced in association with the $\X$ are denoted by ``all''.
Taking the branching faction of the $\X\to J/\psi\pi^+\pi^-$ from the RPP~\cite{ParticleDataGroup:2024cfk}, $(3.5 \pm 0.9) \%$, we get the $\X$ production cross section at the CDF II detector to be
\begin{align}
    \sigma[p\bar p\to X(3872)+\text{all}] = (89\pm30)~\mathrm{nb}.
    \label{eq:xsecCDF}
\end{align}

It was proposed in Ref.~\cite{Bignamini:2009sk} that such a large cross section is in conflict with the hadronic molecular picture of the $\X$, based on the following inequality\footnote{Although here for simplicity we only spell out the $D^0\bar D^{*0}$ component, it should be understood as the proper positive $C$-parity combination of $D^0\bar D^{*0}$ and $\bar D^0D^{*0}$. The same applies to $D\bar D^*$ in the following.}
\begin{align}
    &\quad~ \sigma(\bar{p} p \rightarrow X+\text{all}) \notag\\
    & \sim\left|\int \mathrm{d}^3 \boldsymbol{k}\left\langle X \mid D^0 \bar{D}^{* 0}(\boldsymbol{k})\right\rangle\left\langle D^0 \bar{D}^{* 0}(\boldsymbol{k}) \mid \bar{p} p\right\rangle\right|^2 \notag\\
    & \simeq\left|\int_{\mathcal{R}} \mathrm{d}^3 \boldsymbol{k}\left\langle X \mid D^0 \bar{D}^{* 0}(\boldsymbol{k})\right\rangle\left\langle D^0 \bar{D}^{* 0}(\boldsymbol{k}) \mid \bar{p} p\right\rangle\right|^2 \notag\\
    & \leqslant \int_{\mathcal{R}} \mathrm{d}^3 \boldsymbol{k}|\Psi(\boldsymbol{k})|^2 \int_{\mathcal{R}} \mathrm{d}^3 \boldsymbol{k}\left|\left\langle D^0 \bar{D}^{* 0}(\boldsymbol{k}) \mid \bar{p} p\right\rangle\right|^2 \notag\\
    & \leqslant \int_{\mathcal{R}} \mathrm{d}^3 \boldsymbol{k}\left|\left\langle D^0 \bar{D}^{* 0}(\boldsymbol{k}) \mid \bar{p} p\right\rangle\right|^2,
    \label{eq:inequality}
\end{align}
where the undetected particles are assumed to be spectators, $\mathcal{R}$ is the region of the phase space where the $\X$ is produced, and $\Psi(\boldsymbol{k})$ is the wave function of the $\X$.
It was argued in Ref.~\cite{Bignamini:2009sk} that $\mathcal{R}$ should be of the order of the binding momentum of the $\X$, $\gamma_X\sim 35$~MeV, so that the approximate upper bound in Eq.~\eqref{eq:inequality} is about 0.1~nb, three orders of magnitude smaller than the measured cross section in Eq.~\eqref{eq:xsecCDF}.

The estimate of $\mathcal{R}$ as the binding momentum scale was criticized in Refs.~\cite{Artoisenet:2009wk, Artoisenet:2010uu, Albaladejo:2017blx} (see also the discussion in Ref.~\cite{Guo:2017jvc}).
In Ref.~\cite{Artoisenet:2009wk}, the authors argue that $\mathcal{R}$ should be of the order of the inverse of the force range. 
As a rough estimate, the inverse of the force range is one order of magnitude larger than $\gamma_X$ (see below), and the upper bound in Eq.~\eqref{eq:inequality} is then enlarged by a factor of $\mathcal{O}(10^3)$  in comparison with the estimate in Ref.~\cite{Bignamini:2009sk} and becomes in agreement with the measured value in Eq.~\eqref{eq:xsecCDF}.

We now argue that the inverse of the force range for $D\bar D^*$ $S$-wave interaction should be of the order of a few hundreds of MeV.
Although for the $u$-channel one-pion exchange (OPE) between the $D$ and $D^*$ has a scale of $\mu_\pi = \sqrt{ (M_{D^{*0}}-M_{D^0})^2 - M_{\pi^0}^2}\approx44$~MeV, the main binding force for the $\X$ is likely not the OPE assumed in Ref.~\cite{Tornqvist:1993ng}.
Firstly, renormalization of the OPE potential requires short-distance counter terms (see, e.g., Ref.~\cite{Epelbaum:2008ga}). 
Secondly, phenomenological studies suggest the importance of light vector-meson exchanges (see, e.g., Refs.~\cite{Gamermann:2006nm, Dong:2021juy, Peng:2023lfw, Wang:2023ovj}). 
Thirdly, the binding energies of hidden-charm hadronic molecules predicted in Ref.~\cite{Nieves:2012tta} change only marginally when the OPE is included compared to the case with only constant contact terms.
Fourthly, from analyzing contributions from different Wick contractions in the lattice QCD calculation of the isoscalar $DD^*$ interaction, directly related to the $T_{cc}(3875)^+$, it was found in Ref.~\cite{Chen:2022vpo} that the $\rho$ exchange may be crucial for inducing the attraction between $DD^*$, in line with phenomenological studies~\cite{Dong:2021bvy, Feijoo:2021ppq}; it is reasonable to assume that the $D\bar D^*$ system experiences a similar situation, {which may be tested with lattice QCD calculations}.

\begin{figure}[tb]
    \centering
    \includegraphics[width=0.8\linewidth]{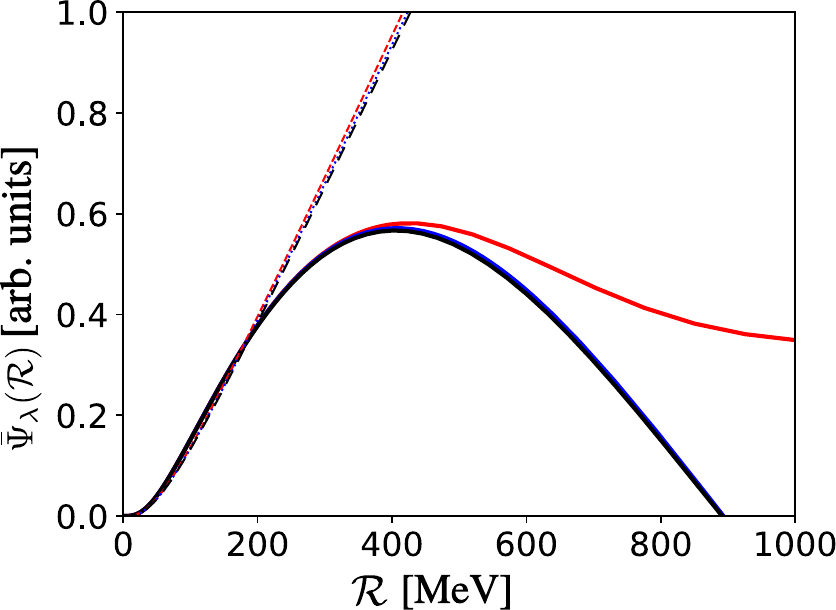}
    \caption{Averaged the wave functions of the deuteron in the $\mathcal{R}$ region. The solid and dashed lines are the results with and without OPE, respectively, and red, blue and black lines correspond to the values of the cutoff $\lambda=0.8$, 1.5 and 4.0~GeV, respectively (the blue and black curves almost overlap with each other). Adapted from Ref.~\cite{Albaladejo:2017blx}. }
    \label{fig:deuteron_wf}
\end{figure}
In order to show that the choice of $\mathcal{R}\simeq \gamma_X$ is too small, in Ref.~\cite{Albaladejo:2017blx}, the authors took the well-understood deuteron as an example. 
The binding energy of the deuteron is about 2.2~MeV, and the binding momentum is $\gamma_d \simeq45$~MeV.
In Fig.~\ref{fig:deuteron_wf}, we show the averaged wave functions of the deuteron in the $\mathcal{R}$ region~\cite{Albaladejo:2017blx}, $\bar{\Psi}_\lambda(\mathcal{R}) \equiv \int_{\mathcal{R}} \mathrm{d}^3 \boldsymbol{k} \Psi_\lambda(\boldsymbol{k})$, where $\lambda$ is the regulator introduced to render the wave function well defined (for details, see Ref.~\cite{Nogga:2005fv}). It is clear that the wave function is far from being saturated by the region $\lesssim\gamma_X$, and the saturation is achieved only when the $\mathcal{R}$ is of a few hundreds of MeV, in line with the argument in Ref.~\cite{Artoisenet:2009wk}.

\begin{figure}[tb]
    \centering
    \includegraphics[width=\linewidth]{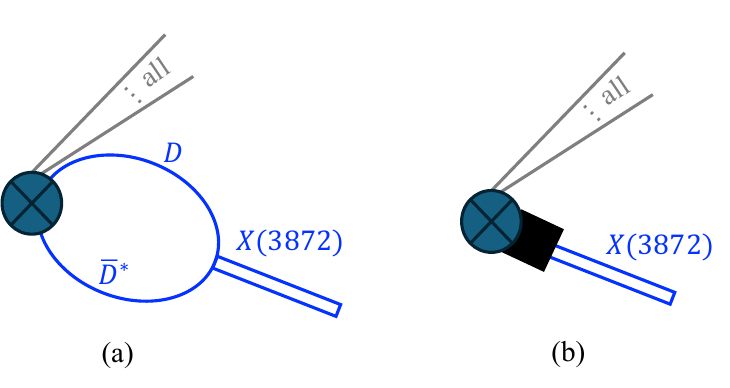}
    \caption{Illustration of the semi-inclusive production of the $\X$ in high-energy collisions. The circled cross presents the source for the $\X$ and its associated particles in the production, denoted as ``all''. (a) is for the $D\bar D^*$ loop contribution, and (b) represents a counter term required to absorb the UV divergence from the loop.}
    \label{fig:semi-inclusive}
\end{figure}
The above point may be understood from considering the diagram corresponding to the second line in Eq.~\eqref{eq:inequality}, shown in Fig.~\ref{fig:semi-inclusive}\,(a).\footnote{Here the charmed meson pair can be both the neutral $D^0\bar D^{*0}$ and the charged $D^+D^{*-}$.}
Since the charmed meson pair is produced at high-energy collisions with all the particles produced in association with the $\X$ assumed to be spectators, the production of $D\bar D^*$ happens at short distances as a point-like vertex. 
Then the loop integral in Fig.~\ref{fig:semi-inclusive}\,(a) is ultraviolet (UV) divergent and can be regularized using a cutoff. A cutoff-dependent counter term is needed to absorb the divergence, as shown in Fig.~\ref{fig:semi-inclusive}\,(b), to render the full amplitude cutoff independent.
For the counter term to be of natural size, that is, of the same order as the loop, the cutoff should also take a natural value for a hard scale, which is again of the order of the inverse of the force range, a few hundreds of MeV. 

The static properties of the $\X$, such as the mass and $J^{PC}$ quantum numbers, are understood through the $D\bar D^*$ hadronic molecular component. However, this does not mean that all its properties are predominantly determined by the long-distance $D\bar D^*$ component. The production of the $\X$ in high-energy collisions is a dynamical process, and the involved mechanism is generally different from that responsible for the internal structure. {It involves the production of a pair of charm and anti-charm quarks at short distances, hadronization of these quarks into charm and anti-charm mesons at intermediate distances ($\sim 1/\Lambda_{\rm QCD}$), and coalescence of these charm mesons into the $\X$ at long distances ($\sim 1/\gamma_X$) (for a discussion of the factorization of long- and short-distance factors for the production and decays of the $\X$, see Ref.~\cite{Braaten:2005jj}).} 
The counter term in Fig.~\ref{fig:semi-inclusive}\,(b) refers to the short-distance production of the $\X$ from sources other than the $D\bar D^*$ intermediate state with low relative momenta, and it includes contributions from, e.g., producing the $\X$ from a $c\bar c$ pair through hadronization.
The prompt production of the $\X$ from $c\bar c$ operators in high-energy hadronic reactions can be studied in the nonrelativistic QCD  framework, see discussions in Refs.~\cite{Artoisenet:2009wk, Meng:2013gga, Butenschoen:2019npa, Cisek:2022uqxa}.

Another way of making the whole production amplitude cutoff independent is to consider the factorization formula as discussed in Ref.~\cite{Braaten:2005jj}. The production of the $\X$, or more generally of hadronic molecules, contains both a long-distance part and a short-distance part. The long-distance part has a typical momentum scale of the $\X$ binding momentum and is given by the $\X$ to $D\bar D^*$ coupling, which is fixed from the binding momentum for a pure composite system~\cite{Weinberg:1965zz, Guo:2017jvc} or equivalently expressed in terms of universal wave function for $S$-wave loosely bound states~\cite{Braaten:2003he, Artoisenet:2009wk}.
The momentum scales in the short-distance part is of the order of the inverse of the force range or larger.
The loop integral in Fig.~\ref{fig:semi-inclusive}\,(a) regularized using a hard three-momentum cutoff is  $G_\Lambda \propto 2\Lambda/\pi + i k$ with $k$ the magnitude of $D$ or $\bar D^*$ in the $D\bar D^*$ c.m. frame.
For $\Lambda$ of the order of the inverse of the force range, the first terms dominates. 
The $\Lambda$ dependence from the loop can be absorbed by the production vertex of the $D\bar D^*$ from high-energy collisions, $P_\Lambda\propto 1/\Lambda$, and the short-distance part of the production amplitude is given by the product $P_\Lambda G_\Lambda$. 

In Refs.~\cite{Guo:2014sca, Albaladejo:2017blx}, the authors made order-of-magnitude estimates of the cross sections for the prompt production in $pp/p\bar p$ collisions of the $\X$ as a $D\bar D^*$ hadronic molecule.
The $D\bar D^*$ cross sections were estimated using Monte Carlo event generators Pythia and Herwig
following Refs.~\cite{Bignamini:2009sk, Artoisenet:2009wk, Artoisenet:2010uu}.
The $D\bar D^*$ loop in Fig.~\ref{fig:semi-inclusive}\,(a) was regularized using a Gaussian form factor with a cutoff $\in[0.5,1.0]$~GeV, corresponding to $\mathcal{R}\in[0.3,0.6]$~GeV~\cite{Albaladejo:2017blx}.
Correspondingly, the estimated cross section for the production of the $\X$ in $p\bar p$ collisions with the CDF kinematics is $\sim [7(5), 29 (20)]$~nb~\cite{Albaladejo:2017blx}, where the values outside (inside) the parentheses were obtained using Herwig (Pythia). The estimate agrees at the order-of-magnitude level with the measured value in Eq.~\eqref{eq:xsecCDF}.
The estimate for the production in $pp$ collisions at 7~TeV is about $\sim[13(4), 55 (15)]$~nb~\cite{Albaladejo:2017blx}, in agreement with $(30\pm9)$~nb\footnote{It was obtained by using 
\[
\begin{aligned}
    \sigma[pp\to X(3872)+\text{all}] &\operatorname{Br}\left[X \rightarrow J / \psi \pi^{+} \pi^{-}\right] \\
    &\approx (1.06 \pm 0.11\pm0.15)~\mathrm{nb},
\end{aligned}
\]
measured by CMS in the kinematic region $10<p_{\mathrm{T}}<30~ \mathrm{GeV}$ and $|y|<1.2$ with $p_T$  and $y$ the transverse momentum and rapidity of the $\X$, respectively~\cite{CMS:2013fpt}.} measured by CMS~\cite{CMS:2013fpt}.

The $\X$ is a hidden-flavor exotic hadron. The dominant component of its wave function for describing its static properties is not necessarily the lowest Fock space components that it can have (which are $q\bar q$ with $q=u,d,s,c$).
In this case, a small $c\bar c$ component may be the driving force for the production of the $\X$ in high-energy collisions even if the $\X$ is predominantly molecular.
Consequently, for a high-energy collision at a c.m. energy squared root of $s$, the scaling (in powers of $s$) of the differential cross section should follow the constituent counting rule~\cite{Lepage:1980fj} for the $c\bar c$ component, despite that its main component still has four constituent (anti)quarks (see also Ref.~\cite{Voloshin:2016phx}).
This is in vast contrast to the deuteron and other light (hyper)nuclei discussion in Section~\ref{sec:exp_HIC1}, whose lowest Fock space component has $3A$ quarks, with $A$ the number of baryons inside the (hyper)nuclei.
Therefore, it is natural that the $\X$ production cross section at large $p_T$ is orders of magnitude larger than those of the deuteron and other light (hyper)nuclei measured by ALICE~\cite{ALICE:2015oer, ALICE:2015wav}, and the critique in Ref.~\cite{Esposito:2015fsa} on the molecular picture of the $\X$ based on the large difference does not hold.

With the same reasoning, since the $f_0(980)$ is also a hidden-flavor meson, the nonobservation of an enhancement of the $p_{{T}}$-differential ${f}_0(980) / {K}^*(892)^0$ ratio in $p$Pb collisions by ALICE~\cite{ALICE:2023cxn} should not be regarded as evidence for the $f_0(980)$ being a normal $q\bar q$ meson or against the $K\bar K$ hadronic molecular picture~\cite{Weinstein:1990gu, Baru:2003qq}.
Similarly, that the scaling of the elliptic anisotropic flow parameter $v_2$ for the production of the $f_0(980)$ in $p$Pb collisions at CMS~\cite{CMS:2023rev} agrees with a two-constituent hypothesis is also expected, and should also not be regarded as evidence for the $f_0(980)$ being a normal $q\bar q$ meson.

\begin{figure}[tb]
    \centering
    \includegraphics[width=\linewidth]{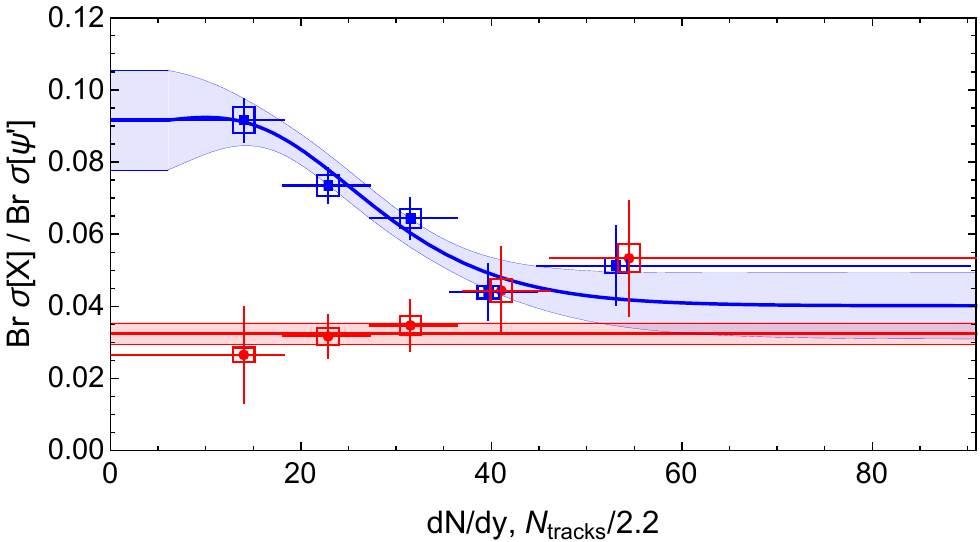}
    \caption{Ratios of the $\X$ and $\psi(2S)$ cross sections (blue: from prompt productions; red: from $b$-hadron decays) measured in the $J / \psi \pi^{+} \pi^{-}$channel as functions of the multiplicity $dN/dy$. 
    The data were measured by LHCb~\cite{LHCb:2020sey}, and the curves with uncertainty bands were obtained from fitting to the LHCb data in Ref.~\cite{Braaten:2020iqw}. From Ref.~\cite{Braaten:2020iqw}.
    }
    \label{fig:multiplicity}
\end{figure}

As discussed in Section~\ref{sec:exp_HIC1}, LHCb measured the productions of the $\X$ and $\psi(2S)$ as functions of the charged-particle multiplicity in $pp$ collisions at 8~TeV~\cite{LHCb:2020sey}.
The measured ratio of the yields of the $\X$ to $\psi(2S)$ in the $J/\psi\pi^+\pi^-$ decay channel decreases as the charged-particle multiplicity increases.
This behavior is in line with the result obtained in the comover interaction model assuming the diameter of the $\X$ (as a compact tetraquark) to be 1.3~fm, and differs drastically from the hadronic molecular picture of the $\X$ obtained with a coalescence model~\cite{Esposito:2020ywk}.
The conclusion was challenged in Ref.~\cite{Braaten:2020iqw}.
In Ref.~\cite{Esposito:2020ywk}, the cross section for breakuping the $\X$ as a hadronic molecule, by scattering with comoving particles, was assumed to be inversely proportional to the $\X$ binding energy.
However, in Ref.~\cite{Braaten:2020iqw}, the authors argued that the breakup cross section can be approximated by the probability-weighted sum of the cross sections for the scattering of the comoving pions with the constituent charmed mesons inside $\X$.
That is because, for the $\X$ with such a small binding momentum $\gamma_X$, the comoving particles can easily probe the internal structure of the $\X$ and thus scatter directly with the constituent particles.
Consequently, the breakup cross section should be insensitive to the $\X$ binding energy.
With a modification of the comover interaction model, the LHCb data can be well described under the hadronic molecular picture of the $\X$, as shown in Fig.~\ref{fig:multiplicity}~\cite{Braaten:2020iqw}.

\subsubsection{The light flavor hadron production}

In previous discussions, exotic hadronic states were primarily examined within systems featuring heavy flavor quarks, such as charm quark and beauty quark. 
There are also many hadronic states in the light flavor sector with properties that do not align with traditional quark model predictions. 
For example, the mass of $\Lambda(1405)$ is lower than that of non-strange $N(1535)$ with the same $J^P$ quantum numbers, even though from the perspective of the three-quark baryon model, $\Lambda(1405)$ contains a heavier $s$ quark. 
In 1950s, the $\Lambda(1405)$ was predicted as a bound state of $K\bar{N}$ by Dalit and Tuan~\cite{Dalitz:1959dn} before its discovery.
Since the 1990s, $\Lambda(1405)$ has been considered as a molecular state of $K\bar{N}$~\cite{Kaiser:1995cy}, accompanied by another nearby pole in the coupled-channel ($\pi\Sigma$-$\bar K N$) scattering amplitudes, known as the two-pole structure for the $\Lambda(1405)$~\cite{Oller:2000fj}. The second and lower pole is now listed as $\Lambda(1380)$  in the latest version of RPP~\cite{ParticleDataGroup:2024cfk}.
Similarly, in the light meson system, the lowest scalar octet are all considered as a ground state family of tetraquarks~\cite{Jaffe:1976ih}, and the $f_0(980)$ is a good candidate of $K\bar K$ molecule~\cite{Weinstein:1990gu, Baru:2003qq}.  

It is also important to investigate possible dibaryon states in order to understand better baryon interactions beyond proton and neutron. So far, the only well-established dibaryon molecular state is the deuteron, while the internal structure of another candidate, $d^*(2380)$~\cite{WASA-at-COSY:2014dmv}, discovered over a decade ago, remains unclear. 
A common explanation is that the $d^*(2380)$ is a double-$\Delta$ state~\cite{Gal:2013dca, Bashkanov:2013cla, Huang:2013nba}. There is also a resonance peak observed below the threshold in the $\bar{K}NN$ system very recently, with preliminary studies indicating $\Lambda N$ as the main decay channel~\cite{J-PARCE15:2024vhg}. 

\begin{figure}[tb]
    \centering
    \includegraphics[width=1.0\linewidth]{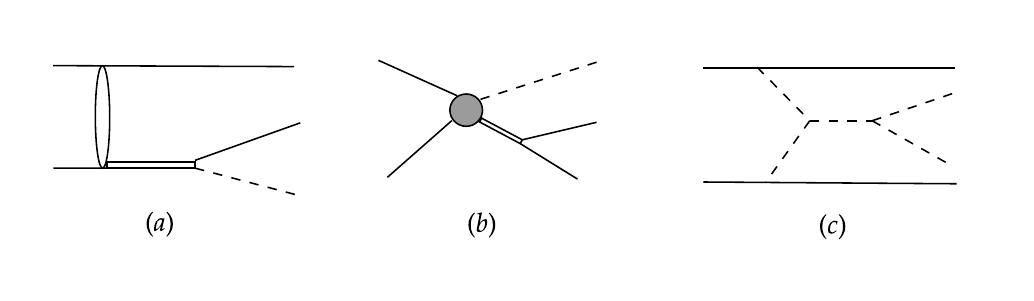}
    \caption{Illustration of the tree diagrams of $NN$ reaction to few-body final states, where the solid lines represent nucleons, dashed lines are for mesons, and double lines represent baryon and dibaryon resonances.}
    \label{fg:diaglight}
\end{figure}

In low-to-moderate energy $pp$ and $p\bar{p}$ collisions, specifically with c.m. energies below 3.5 GeV, especially near the thresholds of certain rare decay channels, exclusive measurements can provide detailed information on many reactions, facilitating sophisticated theoretical modeling. There are huge experimental data samples for $pp$ and $p\bar{p}$ collisions in this energy region from, e.g., IUCF, HADES, CELSIUS, and COSY experiments (for  reviews, see Refs.~\cite{Hanhart:2003pg, Moskal:2002jm}).
Here, we will illustrate how the study of final-state systems in few-body reactions from $pp$ and $p\bar{p}$ collisions can be used to search for baryon excited states, dibaryons, and meson excited states, and discuss  how to identify more exotic states and study their properties.
There are three types of tree diagrams as shown in Fig.~\ref{fg:diaglight}, where (a), (b) and (c) indicate the main mechanism to search for baryon excited states, dibaryons, and meson excited states, respectively.

The search for baryon excited states mainly relies on different final state combinations. 
By applying selection rules based on isospin conservation, regions with significant contributions from certain particles can be identified for further study. 
For example, in the $pp \to pn\pi^+$ process, it is found that although the $N(1440)N\sigma$ has a significant impact~\cite{Ouyang:2008vg}, the largest contribution comes from $\Delta(1232)^{++} \to p\pi^+$, making peaks such as $N(1440)$ less prominent. 
Conversely, in $\bar{p}p \to \bar{p}n\pi^+$, the contribution of $\Delta(1232)^{++}$ is absent, and the cross section for other charged $\Delta(1232)$ states decrease by a factor of 9 due to the isospin factor, leading to increased visibility of peaks such as $N(1440)$~\cite{Wu:2009md}. 
Another example is in the $pp \to n \Sigma^+ K^+$ process~\cite{Tsushima:1998jz, Xie:2007vs}, where $\Sigma^+ K^+$ can only form isospin-3/2 states, providing a good venue for detecting $\Delta$ baryon excited states. 
In previous experiments, measurements have been made for many final states, such as $pp \to pp\pi$, $pp\eta$, $pp \phi$, $pK+\Sigma^0$, $pK+\Lambda$, etc. 
It is important to consider the interaction between initial and final states, with various methods for handling the final state interaction~\cite{Hanhart:2003pg}.
Despite the complexity of these theoretical methods, they can only explain a portion of experimental data. 
For example, in $pp \to pp\eta$, only data for the c.m. energy 12~MeV above the $pp\eta$ threshold can be well described, while the double-peak structure in the $p\eta$ invariant mass spectrum at higher energies cannot be explained by considering only known excited nucleons~\cite{Lu:2015pva}. 
Furthermore, the polarization data cannot be explained by existing models~\cite{Hanhart:2003pg}. 
Another issue is that because of the complication, usually only a single reaction process is analyzed, such as the analysis of  $pp\to pK\Lambda$ in Ref.~\cite{Munzer:2017hbl}. 
However, when considering the $K\Lambda$ resonance states, although the threshold starts from about 1.6~GeV, which covers the $N(1650)$ region,the $N(1535)$ could also have a significant contribution due to the presence of hidden-strangeness in that state~\cite{Liu:2005pm}. A better approach should involve a combined analysis of multiple final states to extract information of multiple resonances. 
Similarly, one can consider to investigate the hidden-charm  $P_c$ states in $pp(\bar{p}) \to pp(\bar{p}) J/\psi$ and $pp(\bar{p}) \to pp(\bar{p}) \eta_c$ reactions~\cite{Wu:2010jy}.

$pp$ and $pd$ collisions offer a good place to search for dibaryon states.
In reactions with either single-pion or double-pion production or with $K\Lambda$/$K\Sigma$ in the final states, apart from the threshold cusp effect in $K\Sigma$, no dibaryon states were observed~\cite{Clement:2016vnl} until 2014 when the Wide Angle Shower Apparatus (WASA) experiment at the Cooler Synchrotron (COSY) established a narrow resonant structure $d^*$ in the $pn \to NN\pi\pi$ reaction, with $I(J^P) = 0(3^+)$ and a width of 70 MeV~\cite{WASA-at-COSY:2014dmv}. 
It was proposed to be a double-$\Delta$ molecular state~\cite{Gal:2013dca, Bashkanov:2013cla, Huang:2013nba} though whether is indeed corresponds to a genuine resonance is still under debate~\cite{Molina:2021bwp}. 
For theoretical predictions prior to the experimental observation, we refer to the review~\cite{Clement:2016vnl}. 
In particular, the quark model prediction of an isospin-3/2 dibaryon in Refs.~\cite{Yuan:1999pg, Dai:2005kt} agrees well with the measured mass. 
A notable feature of the $d^*(2380)$ in theory is that its width is much smaller than that of two $\Delta$ baryons, implying the possible presence of many compact six-quark components in $d^*(2380)$~\cite{Clement:2016vnl, Dai:2023ofz}, although there is still no definitive conclusion. 
This new discovery opens up a new direction in the search for dibaryon states. 
In $pp$ reactions, as shown in Fig.~\ref{fg:diaglight} (b), new dibaryon states may be formed by emitting one or several mesons. 
Recently, in the J-PARC E15 experiment~\cite{J-PARCE15:2024vhg}, a resonance peak structure with a width of 100 MeV was found in the $\Lambda+p$ final state in the reaction $K^- + {^3\rm He} \to (K^- + pp) + n \to (\Lambda+p)+n$ below the threshold of $\bar{K}NN$ by about 40 MeV, indicating the existence of a $\Lambda N$ dibaryon state. 
The existence of this resonance peak can be checked in the $pp\to K \Lambda p$ reaction, but the energy at COSY is just at the threshold for producing this peak structure~\cite{Wilkin:2016mfn}. 
Previous theoretical calculations of this reaction that did not consider the influence of this peak structure need to be revised~\cite{Xie:2011me}. 
For more discussions on the production of dibaryon states in $pp$ collisions, we refer to~\cite{Komarov:2016ebh, Clement:2020dby, Kukulin:2020wvw}.

For the production of excited mesons states, in the low-energy region, the contribution of the diagram shown in Fig.~\ref{fg:diaglight} (c) exchanging two particles is expected to be smaller compared to the other two diagrams. 
However, in the high-energy region, the intermediate exchanged meson can be replaced by a pomeron, and the fusion of two exchanged pomerons to produce mesons plays a crucial role in producing double-pion and double-kaon resonant particles through diffractive processes~\cite{Albrow:2010yb, Lebiedowicz:2018eui}.
Due to the gluon rich environment, this reaction is often used to search for another type of exotic states, glueballs, in high-energy $pp$ or $p\bar{p}$ reactions. 
For calculations on the production of $PC=++$ mesons, such as $f_0(980)$, $f_2(1270)$, $f_0(1500)$, considering the pomeron-pomeron fusion, we refer to, e.g., Refs.~\cite{Szczurek:2009yk, Machado:2011vh, Lebiedowicz:2020bwo}.

Finally, we have to point out that considering solely the tree diagram mechanisms in Fig.~\ref{fg:diaglight} for estimating the production of light hadrons in $pp$ collisions is a very rough approximation. 
In fact, loop contributions in hadronic reactions are often crucial. For reactions with multiple hadrons in the final states, the final-state interactions and coupled-channel effects may be crucial and can significantly affect the extraction of resonance properties. 
Therefore, it is essential to use comprehensive coupled-channel models for few-body systems a combined analysis of various related reactions to extract resonance poles. Tremendous efforts have been made along this directions in, e.g., the J\"ulich-Bonn model~\cite{Ronchen:2012eg}, Argonne-Osaka model~\cite{Matsuyama:2006rp} and three-body unitary models~\cite{Sadasivan:2021emk,Nakamura:2022rdd}. 
However, such a model must include many parameters, and obtaining convincingly determined model parameters requires a significant amount of experimental input. 
More relevant experimental data are desirable.

\subsection{Production in heavy-ion collisions} 
\label{sec:HIC_th}

In this subsection, we will briefly review recent studies of  productions of the $\X$ and $T_{cc}^+$ in heavy-ion collisions. For earlier reviews on related topics, we refer to Refs.~\cite{ExHIC:2013pbl,ExHIC:2017smd,Ohnishi:2016elb,Liu:2024uxn}.

As discussed in Section~\ref{sec:exp_HIC1}, the only signal of the prompt $\X$ production in HIC is from CMS collaboration~\cite{CMS:2021znk}.  
The yields, evolution and distributions of the $\X$, as well as of other exotic hadrons produced in HICs, get complicated by the surrounding QCD medium/nuclear matter/pion gas.

Various coalescence models~\cite{ExHIC:2010gcb,Sun:2017ooe,Fontoura:2019opw,Zhang:2020dwn} and the statistical hadronization model (SHM)~\cite{Andronic:2019wva} were used to estimate the yields of the $\X$ in the hadronic molecular and compact tetraquark pictures. 
The former one depends on a suitable wave function in coordinate space to encode the structure information (for discussions on the subtlety of the short-distance part, see Section~\ref{sec:Xin_pp}). 
In the instantaneous coalescence model, there are still several unclear parameters, for instance the volume size at which coalescence occurs, the available light quark number at the hadronization temperature, and the oscillator frequency of the Wigner function. 
The SHM assumes that hadrons are in a thermal and chemical equilibrium, i.e. with a charm-quark fugacity factor to ensure charm-quark conservation. The yields of the $\X$ in this model only depends on its mass but not on its internal structure. 

\begin{figure}[tb]
\begin{center}
\rotatebox{270}{\includegraphics[width=0.8\linewidth]{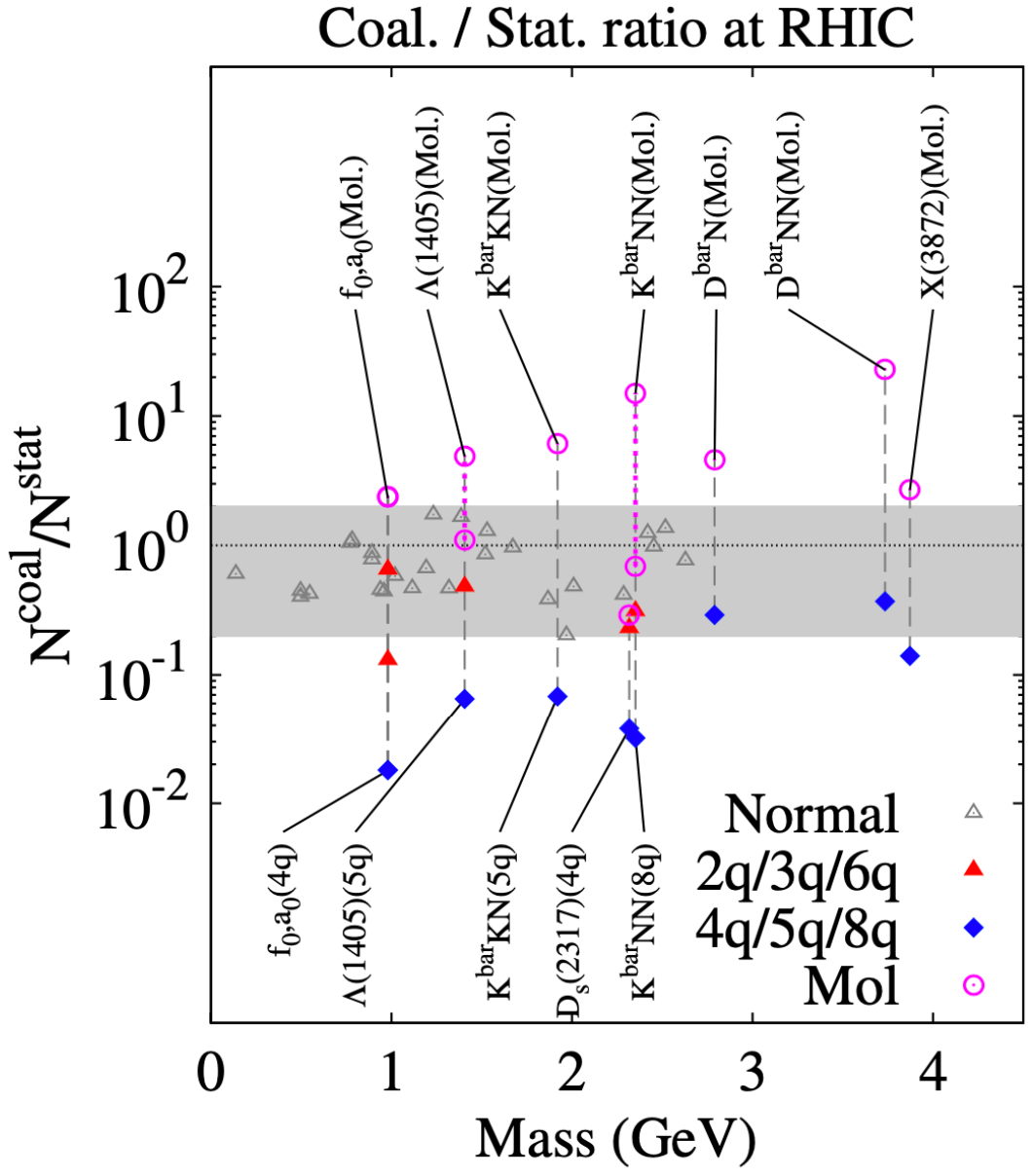}}
\caption{Ratios between the yields in the coalescence model and in the statistic model~\cite{ExHIC:2010gcb}. The gray band is the region for normal hadrons. From Ref.~\cite{ExHIC:2010gcb}.}
\label{fig:ExHIC1}
\end{center}
\end{figure}

In Ref.~\cite{ExHIC:2010gcb}, the authors use both the coalescence model and statistic model to estimate the yields of various hadrons at RHIC, and the model results are summarized in Fig.~\ref{fig:ExHIC1}. 
From the figure, one sees that,
comparing to the yield of normal hadrons, the one of a compact tetraquark is typically one order of magnitude smaller and that of a hadronic molecule is a factor of two or even more.
However, the considered coalescence model does not include the evolution effect in medium,
and the statistic model only depends the masses of hadrons instead of their internal structures. 

\begin{figure}[tb]
\begin{center}
\rotatebox{270}{\includegraphics[width=0.7\linewidth]{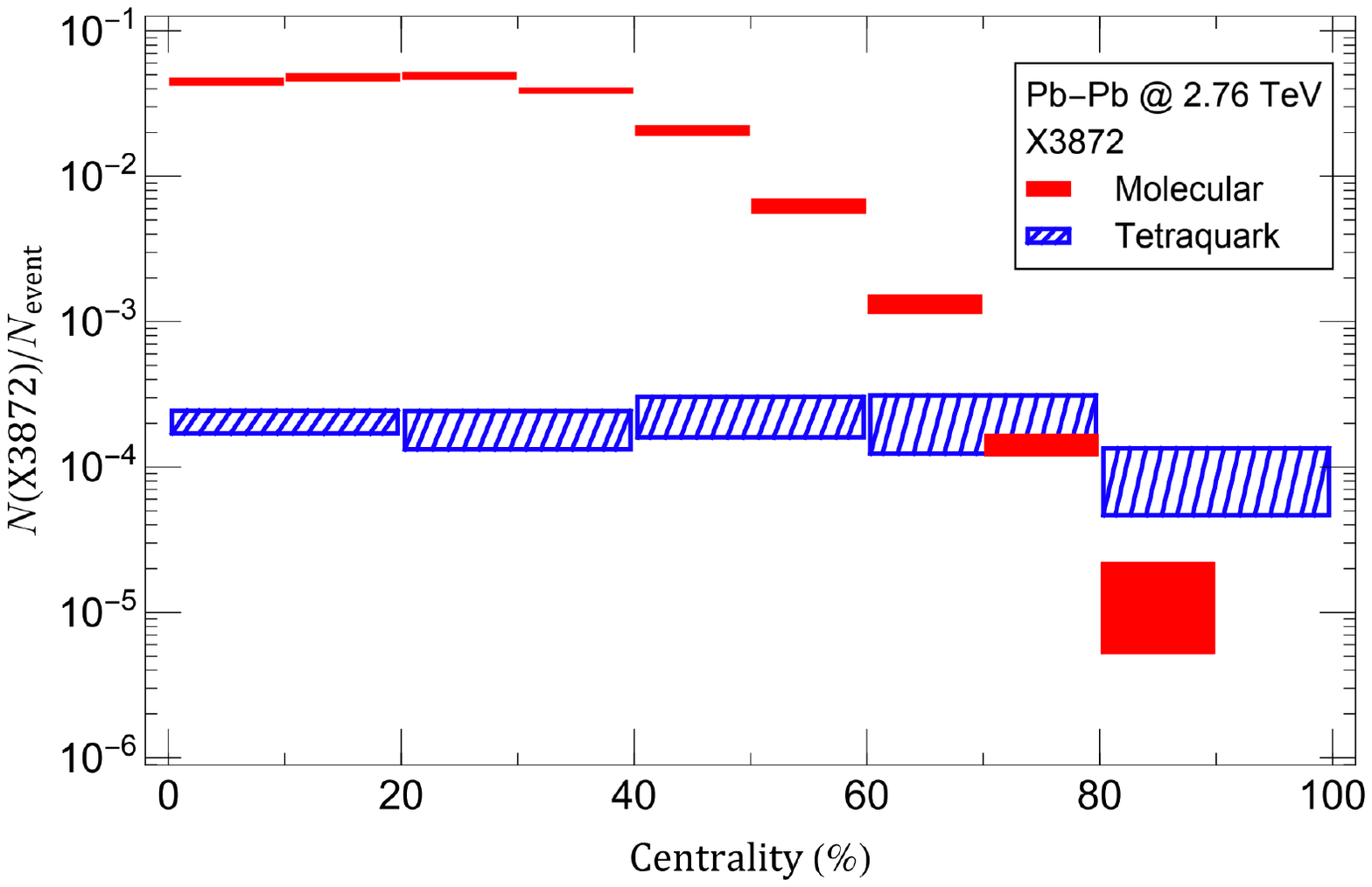}}
\caption{Centrality dependence of the $\X$ in Pb-Pb collisions at $\sqrt{s_{NN}}=2.76~\tev$ in both the molecular (red solid boxes) and compact tetraquark (blue shaded boxes) pictures obtained in Ref.~\cite{Zhang:2020dwn}. The uncertainties are purely statistical. From Ref.~\cite{Zhang:2020dwn}.}
\label{fig:Zhang}
\end{center}
\end{figure}

Further information to deepen the understanding the nature of exotic hadrons is their various distributions,
 for instance the centrality, rapidity and transverse momentum distributions and so on. 
In Ref.~\cite{Zhang:2020dwn}, a multiphase transport model was used to estimate the yield of  the $\X$ in Pb-Pb collisions at $\sqrt{s_{NN}}=2.76~\tev$. 
By considering that the fireball volume is crucial in the production of the $\X$, 
they concluded that the yield of the $\X$ in the molecular picture was two orders of magnitude larger than that of the compact one, and a significant centrality dependence was obtained (see Fig.~\ref{fig:Zhang}). 
In addition, the corresponding rapidity and transverse momentum spectra as well as the elliptic flow coefficient $v_2$ versus transverse momentum for the $\X$ in the molecular picture were also predicted. 
The same method was also applied to estimate the production of double charm tetraquark state $T_{cc}^+$~\cite{Hu:2021gdg} in Pb-Pb collisions at $\sqrt{s_{NN}}=2.76~\tev$ in the molecular picture. 
The ratio between the yield of the $T_{cc}^+$ and that of the $\X$ decreases with the centrality, which agrees with the experimental facts that the yield of the $T_{cc}^+$ is about two orders of magnitude smaller than that of $\X$. 

When the wave function among the constituents is considered, 
for instance the Wigner function included in Ref.~\cite{Chen:2021akx},
the coalescence probability becomes small due to strict constraints on 
the relative momentum between their constituents, 
despite of the large geometric size of hadronic molecules~\cite{Chen:2021akx}. 
In this case, Ref.~\cite{Chen:2021akx} obtains that the total yield of compact tetraquark $\X$  is several times
larger than that of the molecular picture in Pb-Pb collisions. 
At the same time, the fireball volume effect for the centrality dependence 
of the molecular picture is not as significant as that in Ref.~\cite{Zhang:2020dwn}. 
Besides the distributions of the $\X$, the evolution of charm quarks in QGP 
is also included in Ref.~\cite{Chen:2021akx} by the Langevin equation. 
The yields of both scenarios decrease with the evolution time. 
The same method has also been applied to the double charm tetraquark state ~\cite{Chen:2023xhd}.

\begin{figure}[!htbp]
\begin{center}
\rotatebox{270}{\includegraphics[width=0.8\linewidth]{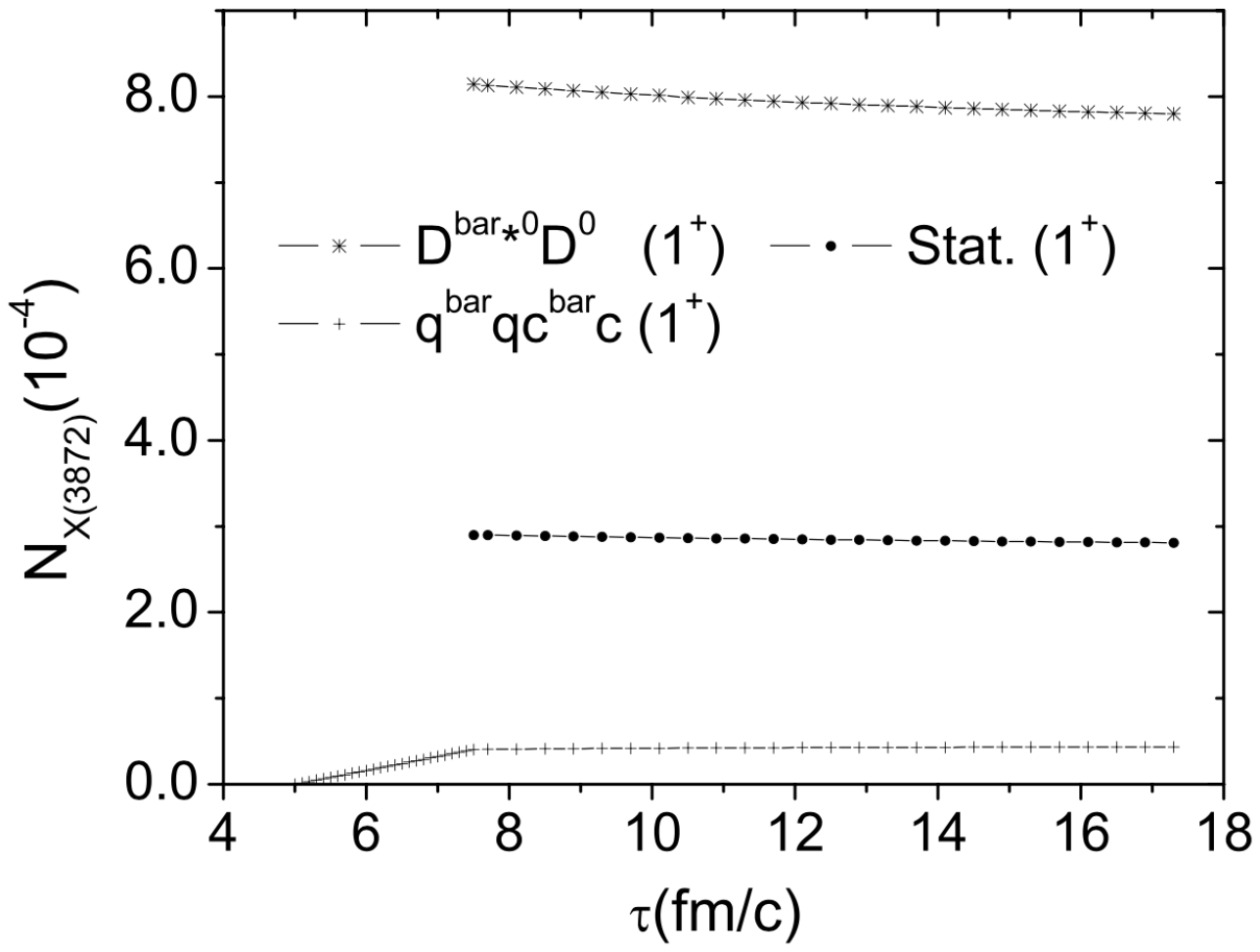}}
\caption{The time evolution of the $\X$ in hadronic molecular and compact tetraquark pictures in central Au-Au collisions at $\sqrt{s_{NN}}=200$~GeV with quantum number $J^P=1^+$ from the calculations in Ref.~\cite{Cho:2013rpa}. From Ref.~\cite{Cho:2013rpa}.}
\label{fig:ExHIC2}
\end{center}
\end{figure}
There is also an early study~\cite{Cho:2013rpa}, based on the SU(4) effective Lagrangians to consider the evolution of the $\X$, by
calculating the corresponding production and absorption cross sections
to estimate the hadronic effects on the $\X$ meson abundance in HICs. 
The absorption cross sections of the $\X$ meson by pions and $\rho$ 
mesons during the hadronic stage of HICs was evaluated.
They estimated the yield of the $\X$ in HICs using both the statistical and coalescence models. 
They found that the absorption cross section was two orders of magnitude larger than the production one and 
the time evolution of the $\X$ abundance in HICs was stable (see Fig.~\ref{fig:ExHIC2}), in contrast to the conclusion of Ref.~\cite{Chen:2021akx}.

The consideration of the production of the $\X$  in HICs is due to 
a quite large number of heavy quarks, as much as 20 $c\bar{c}$ pairs per unit rapidity in Pb-Pb collisions at the LHC energy~\cite{Abreu:2016qci,ExHIC:2010gcb}. 
The charm quarks are free to move through a large volume and can coalescence to form bound states at the end of the QGP phase or the mixture of them. 
A thermally averaged cross section was proposed in Refs.~\cite{Abreu:2016qci,MartinezTorres:2014son}, with 
the dissociation and production processes estimated via $X+\pi\to \bar{D}+D$, $X+\pi\to \bar{D}^*+D^*$ and $D+\bar{D}\to X+\pi$ and $\bar{D}^*+D^*\to \pi+X$, respectively. . 
As those works are based on the coalescence model, their conclusions are similar, 
i.e. the cross sections in the molecular picture are larger than that in the compact tetraquark picture~\cite{Abreu:2016qci,MartinezTorres:2014son}, due to the geometrical arguments~\cite{Abreu:2016qci}. The detailed numbers are scheme dependent. 
However, the geometrical estimate of cross sections cannot be used in high energy collisions, since the particles that scatter with the $\X$ can have enough momentum to probe into the $\X$ and thus the size of the $\X$ is not important any more~\cite{Braaten:2020iqw}.
The same method was also applied to the double charm tetraquark $T_{cc}^+$ in both LHC and RHIC energy region within both molecular and compact tetraquark pictures~\cite{Hong:2018mpk}.

In Ref.~\cite{Wu:2020zbx},  
a transport calculation of the $\X$ trough the fireball formed in Pb-Pb collisions at  $\sqrt{s_{NN}}=5~\mathrm{TeV}$ was performed. 
The formation and dissociation of the $\X$ depend on two transport parameters, 
i.e. its inelastic reaction rate and the thermal equilibrium limit in the evolving hot QCD medium. 
The latter was controlled by the charm production cross section in primordial nucleon-nucleon collisions.
They found that the yield of the loosely bound molecule, assumed to be formed later in the fireball evolution than the tetraquark, was a factor of two smaller than that of the compact tetraquark.  
The same bulk-medium evolution used for charmonium and bottomonium transport was implemented, which is approximated by a cylindrically expanding fireball volume with a transverse flow profile of the blast-wave type, with evolution parameters that reproduce the fits to empirical light-hadron spectra (pions, kaons, protons) at thermal freeze-out temperatures.
This result on the relative yields for the hadronic molecule and compact tetraquark pictures in Ref.~\cite{Wu:2020zbx} is qualitatively different from most coalescence model calculations~\cite{ExHIC:2010gcb,Sun:2017ooe,Fontoura:2019opw,Zhang:2020dwn}; 
In Ref.~\cite{Wu:2020zbx}, the initial hadronic abundance of the molecular configuration is assumed to be zero, which is different 
from the instantaneous coalescence models in Refs.~\cite{ExHIC:2010gcb,Zhang:2020dwn} where the assumption of a wave function provides a large phase space and results in remarkable exceeding in
the equilibrium limit. 
The difference of these results may be regarded as a showcase of the model dependence of estimating the production of exotic hadrons in HICs due to the complication of the QCD medium.

In Ref.~\cite{Cleven:2019cre}, the authors studied the property of the $\X$ in a hot pion bath based on its molecular picture. They find that its width becomes a few tens of MeV at temperature of $100-150~\mev$ and its normal mass moves above the $DD^*$ threshold. 
Their calculation is based on the SU(4) effective Lagrangians and the imaginary time formalism in a self-consistent approach.
The peak associated to the $\X$ becomes significantly wider with increasing temperature. This is because the appearance of a finite imaginary part of the amplitude at the  pole position when temperature effects are included. 
Their result indicates that at typical kinetic freeze out temperatures for RHIC and LHC, the $\X$ cannot be considered as a loosely bound bound state. 
On the contrary, a compact tetraquark-type state, which should couple weakly to $D\bar D^*$ since otherwise a large molecular component is unavoidable, would barely change its behaviour under the same circumstances. 

\begin{figure}[tb]
    \begin{center}
    \rotatebox{270}{\includegraphics[width=0.5\linewidth]{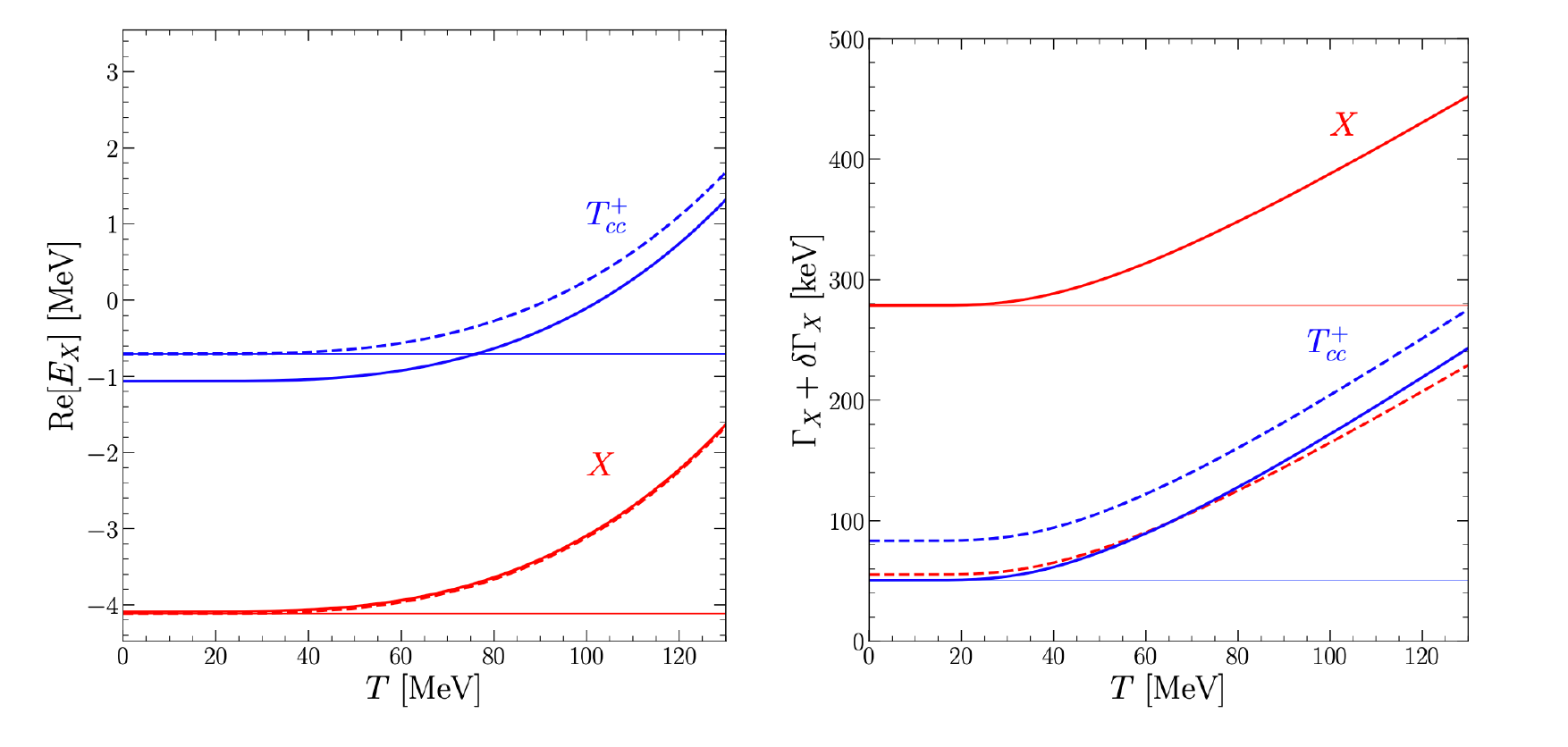}}
    \caption{The real parts of the poles of two-body propagators (left-panel) and the thermal widths (right panel) for the $\X$ (red solid curves), the $T_{cc}^+$  (blue solid curves)  in the pion gas as a function of temperature $T$~\cite{Braaten:2023vgs}. 
    The dashed lines in the left panel are the thresholds of the constituent charm-meson pair, and those in the right panel are the sum of the decays widths of the constituents.
    From Ref.~\cite{Braaten:2023vgs}.}
    \label{fig:X3872_Tcc_Braaten}
    \end{center}
\end{figure}

Recently, a more rigorous treatment of the thermal corrections from the hot pion gas to the propagator of a loosely bound charm-meson molecule was presented using a zero-range effective field theory (ZREFT)~\cite{Braaten:2023vgs}.
One might simply expect that the ZREFT could not be applied to a high temperature, which is characterized by the kinetic freeze-out temperature and is orders of magnitude larger than the binding energy of a loosely bound state. Fortunately, Ref.~\cite{Braaten:2023vgs} illustrates that ZREFT can be applied to such a system by first integrating out thermal pions, leaving the parameters of ZREFT temperature dependent.
The only correction to the binding energy comes from a small temperature-dependent correction to the complex binding momentum. It is noticed that the thermal corrections to the binding energy of the molecule only appear at the next-to-leading order, and the results are shown by Fig.~\ref{fig:X3872_Tcc_Braaten}. These results indicate that the loosely bound molecules, such as the $\X$ and the $T_{cc}^+$, can still survive in the thermal environment of a hadron gas at sufficient low temperatures. This conclusion is consistent with the large rate of the $\X$ production in Pb-Pb collisions by the CMS Collaboration. 

Very recently, in Ref.~\cite{Braaten:2024cke}, the authors proposed that the production rate of a loosely bound hadronic molecule, such as the $\X$, in HICs can be expressed in terms of a short-distance contact density at the kinetic freeze-out of the hadron gas, and it approaches a nonvanishing limit as the binding energy decreases to 0.

Because both the $\X$ and $\psi(2S)$ can be reconstructed in the $J/\psi\pi^+\pi^-$ channel, the production of the $\X $  is often compared with that of the  $\psi(2S)$ (see, e.g., Fig.~\ref{fig:R_Xpsi}). 
For instance, the prompt $\X$ to $\psi(2S)$ yield ratio was found to be~\cite{CMS:2021znk}
\begin{eqnarray}
    \mathcal{R}=\frac{N_{X(3872)}}{N_\psi(2S)}=1.08\pm 0.49(\mathrm{stat})\pm 0.52 (\mathrm{syst})
\end{eqnarray}
with the central value about one order of magnitude higher than the one $0.09$ observed in $pp$ collisions~\cite{LHCb:2020sey}. 
In Ref.~\cite{Abreu:2024mxc}, the authors used a similar method as in Refs.~\cite{Abreu:2016qci,MartinezTorres:2014son} to estimate the thermally averaged cross sections for the production and absorption of the $\psi(2S)$ and used them in the rate equation to determine the time evolution of $N_{\psi(2S)}$. The yield ratio $N_X/N_{\psi(2S)}$ was predicted as a function of the centrality, of the c.m. energy and of the charged hadron multiplicity measured in the mid-rapidity region $[dN_{ch}/d\eta(\eta<0.5)]$. 
\begin{figure}[tb]
    \begin{center}
    \hspace{-0.5cm}\rotatebox{270}{\includegraphics[width=0.7\linewidth]{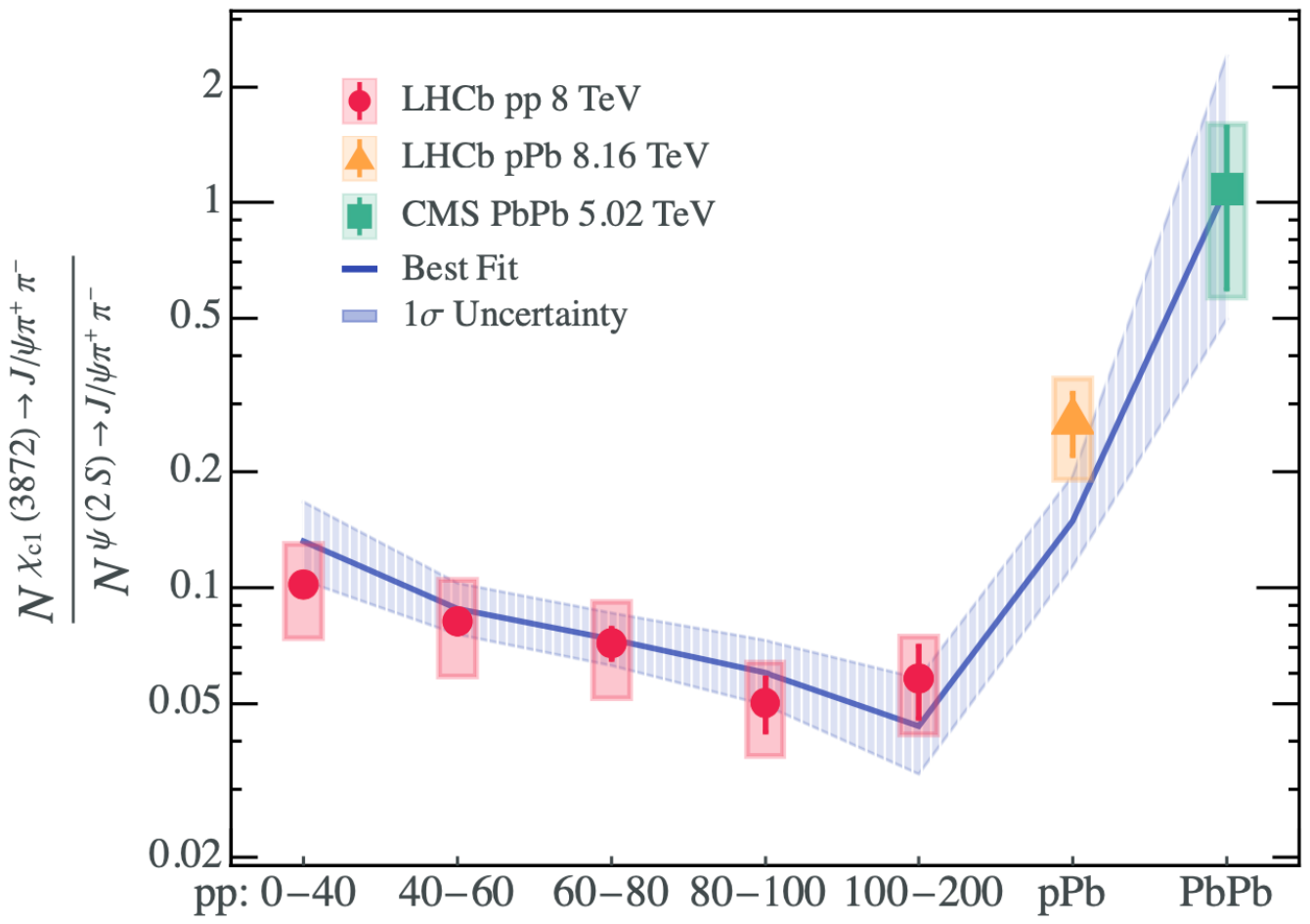}}
    \caption{The yield of the $\X$ relative to that of the $\psi(2S)$ obtained in Ref.~\cite{Guo:2023dwf} in comparison with the LHCb $pp$ collisions data at $\sqrt{s_{NN}}=8~\tev$ (red circle)~\cite{LHCb:2020sey}, LHCb $p$Pb collisions data at $\sqrt{s_{NN}}=8.16~\tev$ (orange triangle)~\cite{LHCb:2022ixc} and CMS Pb-Pb collisions data at $\sqrt{s_{NN}}=5.02~\tev$ (green box)~\cite{CMS:2021znk}. 
    From Ref.~\cite{Guo:2023dwf}.
    }
    \label{fig:X_over_psi2S}
    \end{center}
\end{figure}
Along the same line, Ref.~\cite{Guo:2023dwf} presents a 
phenomenological model for the partonic medium attenuation effects on the $\X$ and $\psi(2S)$ productions in both $pp$ and Pb-Pb collisions. 
A medium-assisted enhancement effect was proposed for the $\X$ production, and it was argued to be dominant at high parton densities and large medium size. Its competition with the absorption-induced suppression leads to a specific pattern of the $N_X/N_{\psi(2S)}$ ratio, which first decreases and then increases 
when the partonic medium evolves from small to large colliding systems. 
A comparison of the results in Ref.~\cite{Guo:2023dwf} with the data is shown by Fig.~\ref{fig:X_over_psi2S}.

{In addition to the measurements of the centrality, transverse momentum, rapidity distributions of exotic hadrons discussed in the previous subsection, one additional important observable that has received intensive interest recently is the momentum correlation between two hadrons. Since the correlation function contains information about the hadron-hadron final state interaction, exotic hadrons that couple to those two hadrons can be studied.}

The momentum correlation function can be expressed in terms of the single particle emission function $S(x_i,\boldsymbol{p}_i)$ which describes the probability of emitting a particle at space-time $x_i$ with momentum $\boldsymbol{p}_i$ through a convolution with the squared relative two-particle wave function,
\begin{eqnarray}\nonumber
    C(\boldsymbol{q}, \boldsymbol{P})=\frac{\int d^4 x_1 d^4 x_2 S_1\left(x_1, \boldsymbol{p}_1\right) S_2\left(x_2, \boldsymbol{p}_2\right)\left|\varphi^{(-)}(\boldsymbol{r}, \boldsymbol{q})\right|^2}{\int d^4 x_1 S_1\left(x_1, \boldsymbol{p}_1\right) \int d^4 x_2 S_2\left(x_2, \boldsymbol{p}_2\right)}.
\end{eqnarray}
The relative wave function $\left|\varphi^{(-)}(\boldsymbol{r}, \boldsymbol{q})\right|^2$ is for the hadron pair in the outgoing state that contains information about the hadron-hadron interaction. 
If the two hadrons are totally independent with each other, the momentum correlation functions should equal to unity. 
The correlation function is often approximated as (see, e.g., Ref.~\cite{Lisa:2005dd})
\begin{eqnarray}
    C(\boldsymbol{q}, \boldsymbol{P})=\int d^3 \boldsymbol{r} S_{12}(\boldsymbol{r})\left|\varphi^{(-)}(\boldsymbol{q}, \boldsymbol{r})\right|^2.
\end{eqnarray}
Usually, a spherical Gaussian source function with a given source radius is assumed, which ignores the dynamical property of the particle emission sources.
Low-energy scattering observables, including scattering lengths and effective ranges, for various systems were extracted with this method. 
Lots of work has been done along this line, see, e.g., Refs.~\cite{ExHIC:2017smd,Ohnishi:2016elb,ExHIC:2013pbl,Morita:2019rph}; for a recent review, see Ref.~\cite{Liu:2024uxn}.
However, one should notice that results on hadron-hadron interactions extracted using this method depend on the assumption of the source profile and the source size.
While the low-energy scattering observables correspond to long-distance physics, the source profile is of short-distance nature.
On the one hand, the source function serves as a UV regulator (a form factor) for the production of the hadron pair, and the source size acts as a cutoff in the form factor.
On the other hand, the correlation function measured in experiments, as physical observables, should not depend on the UV regulator.
From this point of view, works still need to be done in order to get the hadron-hadron interactions in a model-independent way.
{It is worthwhile to notice that recent studies by ALICE~\cite{ALICE:2020ibs, ALICE:2023sjd}, which modeled the source considering a Gaussian profile and an exponential tail due to strongly decayed resonances, suggested a common source for meson-meson and meson-baryon sectors in $pp$ collisions at the LHC. }

\section{Summary and Outlook}

 Exotic hadrons remain a vibrant area of research, with significant contributions from various experiments, including Belle, BaBar, BESIII, LHC experiments, RHIC, etc. These experiments provide a broad and detailed understanding of exotic states, pushing the boundaries of our knowledge of QCD and the strong interaction. 
 Here we have presented a concise review of the studies of exotic hadrons in $pp/p\bar p$ and nuclear collisions.
 More and more data are being collected on the prompt production in such collisions. However, from the discission above, one sees that more efforts are still needed to understand the production mechanisms of exotic hadrons and to shed lights into nature of exotic hadrons.

 {Last but not at least, let us mention that the strucure of exotic hadrons need to be understood with a combination of different reactions that provide supplementary information. In addition to the nuclear collisions reviewed here, $e^+e^-$ collisions, $b$-flavored hadron decays and photoproductions also play their unique roles in studying exotic hadrons. For instance, $\X\gamma$~\cite{BESIII:2013fnz} and $Z_c(3900)$~\cite{BESIII:2013ris, Belle:2013yex} were observed in $e^+e^-$ collisions only at a specific range of energies around the mass of the $\psi(4230)$. This feature could hint at the production mechanism for these states to be due to charmed-meson intermediate states that couple strongly to these exotic particles, and thus provide invaluable input to understanding them. }

 Future upgrades of currently running experiments and ongoing research {using various reactions} promise further discoveries and insights into the nature of exotic hadrons.
\section*{Acknowledgments}

We are grateful to Baoyi Chen, Xingyu Guo, and Shuze Shi for  useful discussions.
This work is supported in part by the National Key Research and Development Program of China under Contract Nos. 2022YFA1604900 and 2023YFA1606703; by the National Natural Science Foundation of China under Grant Nos.~12025501, 12147101, 12375073, 12125507, 12361141819, 12047503, 12175239, and 12221005; and by Chinese Academy of Sciences under Grant Nos. XDB34000000 and YSBR-101.

\bibliographystyle{woc.bst}
\bibliography{refs}

\end{document}